\documentclass[ALICE,manyauthors]{cernphprep}

\usepackage{xcolor}
\usepackage[comma,square,numbers,sort&compress]{natbib}
\usepackage{lineno}
\usepackage{multirow}
\usepackage{hyperref} 

\newcommand{\Eq}[1]   {Eq.~(\ref{#1})}

\newcommand{\Fi}[1]   {Fig.~\ref{#1}}
\newcommand{\Fis}[2]  {Figs.~\ref{#1} and~\ref{#2}}

\newcommand{\Ta}[1]   {Tab.~\ref{#1}}

\newcommand{\gevc}    {\mbox{GeV$/c$}}
\newcommand{\gevcsq}  {\mbox{GeV$^{\:2}/c^{2}$}}
\newcommand{\gevcc}   {\mbox{GeV$/c^2$}}

\newcommand{\invmub}  {\mbox{$\mu$b$^{-1}$}}
\newcommand{\rb}[1]   {\mbox{\textrm{\scriptsize #1}}}
\newcommand{\rbt}[1]  {\mbox{\textrm{\tiny #1}}}
\newcommand{\jpsi}    {\ensuremath{\textrm{J}/\psi}}

\newcommand{\chic}    {\ensuremath{\chi_{c}}}
\newcommand{\psitwos} {\ensuremath{\psi(\textrm{2S})}}

\newcommand{\epem}    {\ensuremath{\textrm{e}^{+} \textrm{e}^{-}}}

\newcommand{\ccbar}   {\ensuremath{\textrm{c}\bar{\textrm{c}}}}
\newcommand{\ppbar}   {\ensuremath{\textrm{p}\bar{\textrm{p}}}}
\newcommand{\sqrts}   {\ensuremath{\sqrt{s}}}
\newcommand{\sqrtsnn} {\ensuremath{\sqrt{s_{_{\rbt{NN}}}}}}
\newcommand{\pt}      {\ensuremath{p_{\rb{T}}}}
\newcommand{\zt}      {\ensuremath{z_{\rb{T}}}}
\newcommand{\ptavg}   {\ensuremath{\langle p_{\rb{T}} \rangle}}
\newcommand{\ptavgsq} {\ensuremath{\langle p_{\rb{T}}^{2} \rangle}}

\newcommand{\minv}    {\ensuremath{m_{\rb{ee}}}}

\newcommand{\dedx}    {\ensuremath{\textrm{d}E/\textrm{d}x}}

\newcommand{\npart}   {\ensuremath{\langle N_{\rb{part}} \rangle}}
\newcommand{\ncoll}   {\ensuremath{\langle N_{\rb{coll}} \rangle}}
\newcommand{\taa}     {\ensuremath{\langle T_{\rbt{AA}} \rangle}}

\newcommand{\der}     {\ensuremath{\textrm{d}}}

\newcommand{\chisq}   {\ensuremath{\chi^{2}}}

\newcommand{\raa}     {\ensuremath{R_{\rbt{AA}}}}
\newcommand{\ptsqraa} {\ensuremath{r_{\rbt{AA}}}}
\newcommand{\lumint}  {\ensuremath{{\cal L}_{\rb{int}}}}
\newcommand{\acceff}  {\ensuremath{{A \times \epsilon}}}
\newcommand{\fb}      {\ensuremath{f_{\rbt{B}}}}
\newcommand{\fbprime} {\ensuremath{f_{\rbt{B}}^{\prime}}}
\newcommand{\fsig}    {\ensuremath{f_{\rbt{Sig}}}}
\newcommand{\ffb}     {\ensuremath{F_{\rbt{Bkg}}}}
\newcommand{\ffs}     {\ensuremath{F_{\rbt{Sig}}}}
\begin{document}
\begin{titlepage}
\PHyear{2015}
\PHnumber{092}      
\PHdate{1 April}     
\title{Inclusive, prompt and non-prompt \jpsi\ production at
  mid-rapidity in Pb-Pb collisions at \sqrtsnn~= 2.76~TeV}
\ShortTitle{Inclusive, prompt and non-prompt \jpsi\ production in
  Pb-Pb collisions}
\Collaboration{ALICE Collaboration\thanks{See
    Appendix~\ref{app:collab} for the list of collaboration members}}
\ShortAuthor{ALICE Collaboration} 

%
\begin{abstract}
The transverse momentum (\pt) dependence of the nuclear modification
factor \raa\ and the centrality dependence of the average transverse
momentum \ptavg\ for inclusive \jpsi\ have been measured with ALICE
for Pb-Pb collisions at \sqrtsnn~= 2.76~TeV in the \epem~decay channel
at mid-rapidity ($|y| < 0.8$).  The \ptavg\ is significantly smaller
than the one observed for pp collisions at the same centre-of-mass
energy.  Consistently, an increase of \raa\ is observed towards low
\pt.  These observations might be indicative of a sizable contribution
of charm quark coalescence to the \jpsi\ production.  Additionally,
the fraction of non-prompt \jpsi\ from beauty hadron decays, \fb, has
been determined in the region $1.5 < \pt < 10$~\gevc\ in three
centrality intervals.  No significant centrality dependence of \fb\ is
observed.  Finally, the \raa\ of non-prompt \jpsi\ is discussed and
compared with model predictions.  The nuclear modification in the
region $4.5 < \pt < 10$~\gevc\ is found to be stronger than predicted
by most models.
\end{abstract}
\vspace{2cm}
%
%
\end{titlepage}
\setcounter{page}{2}

%
\section{Introduction}

Heavy-ion collisions at high energies allow the study of strongly
interacting matter under extreme conditions.  Calculations based on
Quantum-Chromo-Dynamics (QCD) on the lattice indicate that the hot and
dense medium created in these collisions behaves like a strongly
coupled Quark-Gluon Plasma (QGP)
\cite{Karsch:2006xs,Borsanyi:2010bp,Borsanyi:2010cj,Bazavov:2011nk}.
Heavy quarks are an important probe for the properties of this state
of matter, since they are produced via hard partonic collisions at a
very early stage and thus experience the complete evolution of the
system.  Quarkonium states, i.e. bound states of a heavy quark and
anti-quark such as the \jpsi~meson (\ccbar~state) are of particular
interest.  It was predicted that the \jpsi~formation is suppressed in
a QGP due to the screening of the \ccbar~potential in the presence of
free colour charges \cite{Matsui:1986dk}.  Experimentally, a
suppression of the inclusive \jpsi~yield in heavy-ion collisions
relative to the corresponding yield in pp, scaled by the number of
binary nucleon-nucleon collisions, has been observed at the Super
Proton Synchrotron (SPS)
\cite{Alessandro:2004ap,Arnaldi:2009ph,Brambilla:2010cs} and the
Relativistic Heavy Ion Collider (RHIC)
\cite{Adare:2006ns,Adare:2011yf}.  The level of suppression was found
to be similar at SPS and RHIC, despite the significantly different
collision energy.  More recently, the nuclear modification of \jpsi\
was also measured for Pb-Pb collisions at the LHC
\cite{Abelev:2012rv,Abelev:2013ila,Chatrchyan:2012np}.  While at high
transverse momentum ($\pt > 4$~\gevc) the suppression factor is at the
same level as the one observed at RHIC in the low \pt~region, a
significant reduction of the suppression is measured towards lower
\pt.  This has been interpreted as the effect of an additional
contribution to \jpsi\ production at low \pt, due to the combination
of correlated or uncorrelated $\textrm{c}$ and $\bar{\textrm{c}}$
quarks \cite{BraunMunzinger:2000px,Thews:2000rj}.  This contribution
becomes sizable at LHC energies, since the number of \ccbar~pairs is
much higher than at lower energies.  Assuming that a deconfined phase
is produced and that all the \jpsi\ are dissociated, this process
happens at the chemical freeze-out stage of the fireball evolution.
This is the approach followed within the statistical hadronization
models described in Refs.~\cite{Gorenstein:2000ck,Andronic:2003zv}.
Alternatively, \jpsi\ could be generated via coalescence throughout
the full evolution of the QGP phase, if their survival probability in
this environment is large enough.  This scenario has been implemented
in several partonic transport models \cite{Zhao:2007hh,Liu:2009nb}.
It was found that both approaches can provide a description of the
measured nuclear modification factors \cite{Abelev:2013ila} and of the
elliptic flow of inclusive \jpsi\ \cite{ALICE:2013xna}. 

The production of open beauty hadrons is expected to be sensitive to
the density of the medium created in heavy-ion collisions due to the
energy loss experienced by the parent parton (a beauty quark) which
hadronizes into the beauty hadron.  This energy loss is expected to
occur via medium-induced gluon radiation
\cite{Gyulassy:1990ye,Baier:1996sk} and elastic collisional energy
loss processes \cite{Thoma:1990fm,Braaten:1991jj,Braaten:1991we} and
it depends on the QCD Casimir coupling factor of the parton (larger
for gluons than for quarks) and on the parton mass
\cite{Dokshitzer:2001zm,Armesto:2003jh,Wicks:2007am,Zhang:2003wk}.
Other mechanisms, such as in-medium hadron formation and dissociation,
can be envisaged as particularly relevant for heavy-flavour hadrons
due to their small formation times
\cite{Sharma:2009hn,Sharma:2012dy,Adil:2006ra}.

Inclusive \jpsi\ production is the sum of several contributions.  In
addition to the directly produced \jpsi, the decays of heavier
charmonium states, such as the \chic\ and \psitwos, also contribute to
the inclusive \jpsi\ yield.  These two sources (direct and charmonium
decays) are defined as prompt \jpsi, where the contribution from
charmonium decays is about 35\% as measured in pp collisions
\cite{Faccioli:2008ir}.  Since heavier charmonia are less strongly
bound than the \jpsi\ they should be more easily dissolved in a
deconfined medium \cite{Mocsy:2007jz}.  The \jpsi\ suppression
measured at the SPS is indeed compatible with the assumption that only
the excited states are dissolved and not the directly produced \jpsi\
\cite{Alessandro:2004ap,Arnaldi:2009ph}.  On top of the prompt \jpsi\
production, there is an additional non-prompt contribution to the
inclusive \jpsi\ at high centre-of-mass energies, coming from the
decay of beauty hadrons.  Since these decays proceed via weak
interactions, the resulting \jpsi\ will originate from a decay vertex
that is displaced from the main interaction vertex.  Their measurement
provides a direct determination of the nuclear modification of beauty
hadrons.  By subtracting the non-prompt contribution from the
inclusive \jpsi\ yield one can also provide an unbiased information on
medium modification of prompt charmonia.  The non-prompt
\jpsi~contribution at mid-rapidity has already been measured in
pp~collisions at \sqrts~= 7~TeV by ATLAS \cite{Aad:2011sp}, CMS
\cite{Chatrchyan:2011kc} and ALICE \cite{Abelev:2012gx}.  For Pb-Pb
collisions at \sqrtsnn~= 2.76~TeV CMS has also published prompt and
non-prompt \jpsi\ production results at mid-rapidity for $\pt >
6.5$~\gevc\ \cite{Chatrchyan:2012np}.

In this paper we present a differential measurement of the inclusive
\jpsi\ production at mid-rapidity in Pb-Pb collisions at \sqrtsnn~=
2.76~TeV.  The \pt\ dependence of the nuclear modification factor and
the centrality dependence of the average transverse momentum of \jpsi\
have been obtained, extending the set of results presented in
\cite{Abelev:2013ila}.  A measurement of the prompt and non-prompt
contributions to the inclusive \jpsi\ production is also presented.
The nuclear modification factor of non-prompt \jpsi\ is determined
down to \pt~= 1.5~\gevc\ and compared to model predictions.

%
\section{Data Analysis}

A detailed description of the ALICE detector can be found in
\cite{Aamodt:2008zz}.  For the analysis presented here the detectors
of the central barrel have been used, in particular the Inner Tracking
System (ITS) and the Time Projection Chamber (TPC).  These detectors
are located inside a large solenoidal magnet with a field strength of
0.5~T.  They allow the measurement of \jpsi~mesons via the dielectron
decay channel in the central rapidity region down to zero \pt.  The
ITS \cite{Aamodt:2010aa} consists of six layers of silicon detectors
surrounding the beam pipe at radial positions between 3.9~cm and
43.0~cm.  Its two innermost layers are composed of Silicon Pixel
Detectors (SPD), which provide the spatial resolution to separate on a
statistical basis the non-prompt \jpsi.  The active volume of the TPC
\cite{Alme:2010ke} covers the range along the beam direction $-250 <
z < 250$~cm relative to the Interaction Point (IP) and extends in
radial direction from 85~cm to 247~cm.  It is the main tracking device
in the central barrel and is in addition used for particle
identification via the measurement of the specific ionization (\dedx)
in the detector gas.

Triggering and event characterization is performed via forward
detectors, the V0 \cite{Abbas:2013taa} and two Zero Degree
Calorimeters (ZDC) \cite{ALICE:2012aa}.  The V0 detectors consist
of two scintillator arrays positioned at $z = -90$~cm and $z =
+340$~cm and cover the pseudo-rapidity ranges $-3.7 \le \eta \le -1.7$
and $2.8 \le \eta \le 5.1$.  The ZDCs, each one consisting of two
quartz fiber sampling calorimeters, are placed at a distance of 114~m
relative to the IP in both directions along the beam axis and are used
to detect spectator nucleons.

The results presented in this article are based on data samples
collected during the Pb-Pb data taking periods of the LHC in the years
2010 and 2011.  In the case of the 2011 data sample the Minimum Bias
(MB) Level-0 (L0) trigger condition was defined by the coincidence of
signals in both V0 detectors along with a valid bunch crossing
trigger.  For the 2010 data sample, in addition, the detection of at
least two hits in the ITS was required.  Both MB trigger definitions
lead to trigger efficiencies larger than 95\% for inelastic Pb-Pb
collisions.  Electromagnetic interactions were rejected by the Level-1
(L1) trigger, which required a minimum energy deposition in the ZDC by
spectator neutrons.  The beam-induced background was further reduced
during the offline analysis by selecting events according to the
relative timing of signals in V0 and ZDC.  The offline centrality
selection is done using the sum of the two V0 signal amplitudes.
By fitting the corresponding distribution with the results of Glauber
model simulations, the average number of participants \npart\ and the
average nuclear overlap function $\taa = \ncoll /
\sigma_{\rbt{NN}}^{\rb{inel}}$ for a given centrality class can be
determined as described in \cite{Abelev:2013qoq}.  Here, \ncoll\ is
the average number of binary nucleon-nucleon collisions and
$\sigma_{\rbt{NN}}^{\rb{inel}}$ the inelastic nucleon-nucleon cross
section.  The numerical values for \npart, \ncoll, and \taa\ are
tabulated in \cite{Abelev:2013ila}.

The 2010 data sample consists of $1.5 \times 10^{7}$ events, taken
with the corresponding MB trigger.  The 2011 event sample was enriched
with central and semi-central Pb-Pb collisions by using thresholds on
the V0 multiplicity at the L0 trigger.  From the latter data set we
analyzed $1.9 \times 10^{7}$ central ($0$--$10$\% of the centrality
distribution) and $1.7 \times 10^{7}$ semi-central ($10$--$50$\%)
events.  The summed 2010 and 2011 data samples correspond to an
integrated luminosity of $\lumint = 26.4 \pm 0.3\textrm{(stat.)}
^{+2.1}_{-1.7}\textrm{(syst.)}$~\invmub\ \cite{Abelev:2013ila}.

%
\subsection{Inclusive \jpsi}
\label{sec:incljpsirec}

\jpsi\ candidates are reconstructed by combining opposite-sign (OS)
pairs of electron/positron candidates and calculating their invariant
mass \minv.  These candidates are selected from tracks reconstructed
in the ITS and the TPC by employing the set of quality criteria
described in \cite{Abelev:2013ila,Aamodt:2011gj}.  In order to reject
the background from photon conversions in the detector material,
tracks are required to have a hit in one of the SPD layers.  In
addition, at least 70 out of a maximum of 159 space points
reconstructed in the TPC must be assigned to a given track, which also
needs to fulfill a quality criterion of the track fit ($\chisq/ndf <
4$).  The tracks are required to be in the range $|\eta| < 0.8$, where
the tracking and particle identification performance of the TPC is
optimal, and to have $\pt > 0.85$~\gevc\ to improve the
signal-to-background ratio in the \jpsi\ mass region.

Electron candidates are selected by requiring that the \dedx\
measurement in the TPC lies within a band $[-1 \sigma, +3\sigma]$ 
around the momentum-dependent parameterization of the expected
signal, where $\sigma$ is the phase space dependent \dedx\ resolution
(details can be found in \cite{Abelev:2014ffa}).  The selection is
asymmetric in order to minimize the contribution from pions.  To
further suppress the hadron contamination, tracks that are compatible
within $\pm 4 \sigma$ with the proton expectation are rejected.  A
side effect of this cut is that tracks below $\pt = 1$~\gevc\ are
effectively removed.

%
\begin{figure}
\begin{center}
\begin{minipage}[b]{0.49\linewidth}
\includegraphics[width=1.0\linewidth]{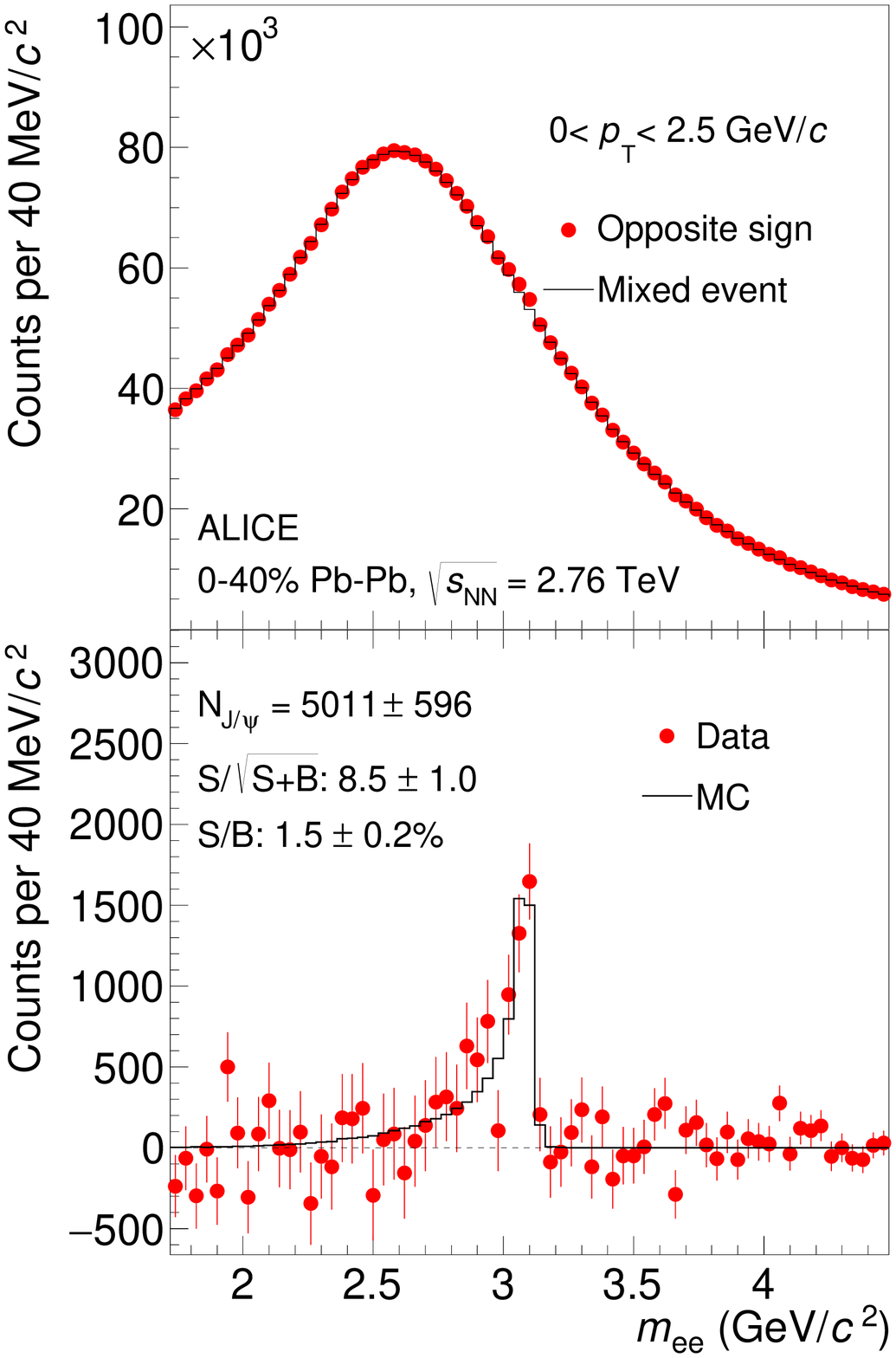}
\end{minipage}
\begin{minipage}[b]{0.49\linewidth}
\includegraphics[width=1.0\linewidth]{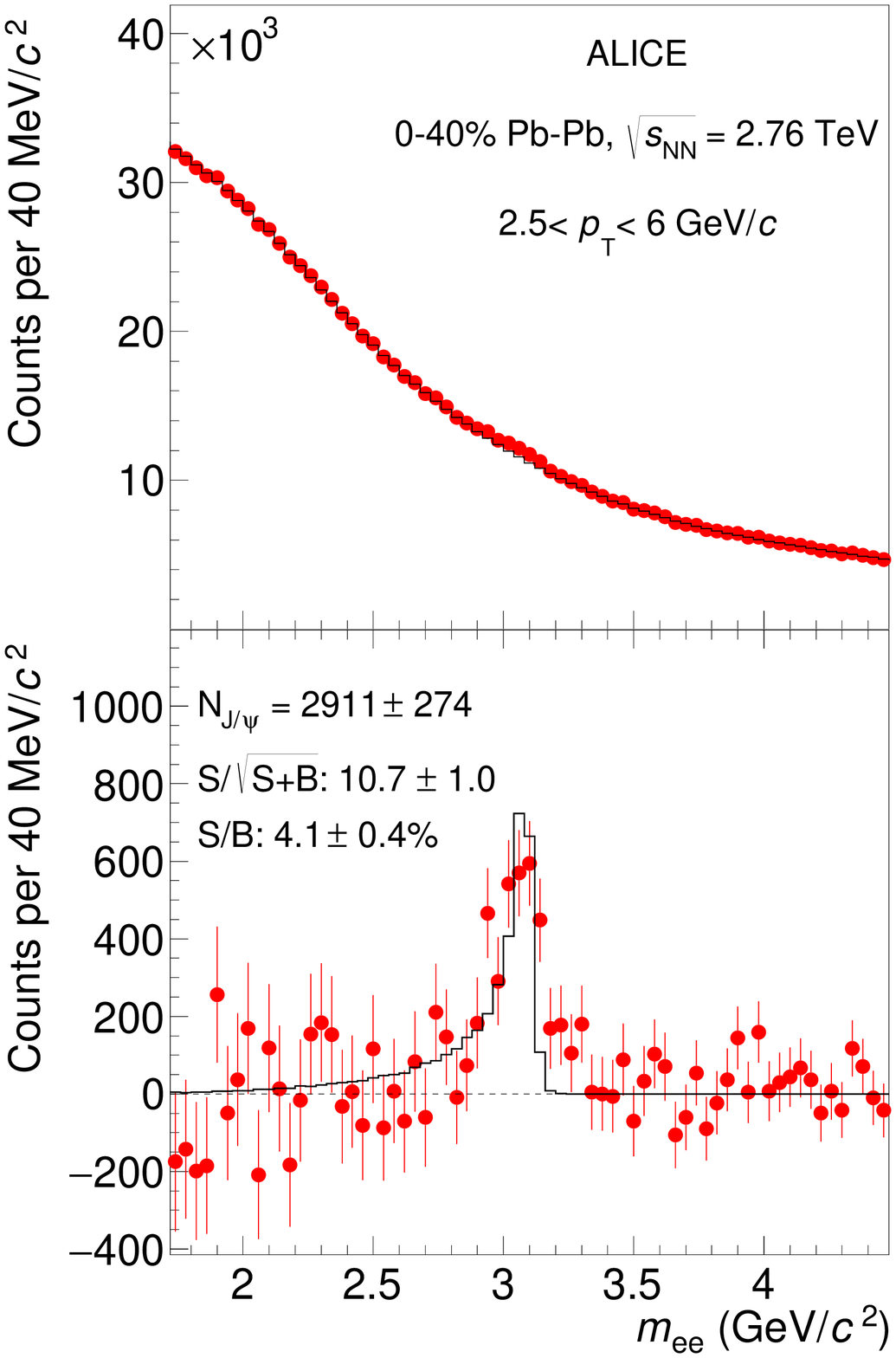}
\end{minipage}
\end{center}
\caption{\label{fig:minvincl}
The invariant mass distributions of inclusive \jpsi\ at mid-rapidity
($|y| < 0.8$) for Pb-Pb collisions ($0$--$40$\% most central) at
\sqrtsnn~= 2.76~TeV.  The left panels show the interval $0 < \pt <
2.5$~\gevc\ and the right ones $2.5 < \pt < 6$~\gevc.  The upper
panels display the opposite sign distributions together with the result
of the mixed event procedure.  In the lower panels the background
subtracted distributions are shown and compared to the simulated line
shape.  Also, the signal-to-background ratio $S/B$ and the
significance of the signal are given.}
\end{figure}
%

{\it Measurement of the inclusive \jpsi\ yield} \\
The \jpsi\ signal counts $N_{\jpsi}$ are obtained from the number of
entries in the background subtracted invariant mass distributions in
the range $2.92 < \minv < 3.16$~\gevcc.  The uncorrelated background
is evaluated with a mixed event (ME) technique.  In order to achieve
a good description of the background only electrons and positrons from
events with similar properties in terms of centrality, primary vertex
position, and event plane angle are combined.  The ME distributions
are scaled to the same event (SE) distributions in the mass ranges
$1.5 < \minv < 2.5$~\gevcc\ and $3.2 < \minv < 4.2$~\gevcc, so that
the \jpsi\ signal region is excluded.  The normalization area contains
the \psitwos\ signal, but its contribution is negligible and can
therefore be safely ignored.  Also, contributions from the tail of the
\jpsi\ signal shape to this mass interval are below the percent level
and will thus not significantly affect the normalization.
Figure~\ref{fig:minvincl} shows a comparison of the SE and ME
invariant mass distributions for the $0$--$40$\% most central Pb-Pb
collisions for electron-positron pairs at mid-rapidity ($|y| < 0.8$)
in two \pt~intervals: 0~--~2.5~\gevc\ and 2.5~--~6~\gevc.  The
agreement between the SE and ME distributions outside the signal
region is very good and allows signal extraction with significances
larger than eight.

The \jpsi~yield per MB event in a given \pt~interval, $Y_{\jpsi}$, is
obtained as
\begin{equation}
  Y_{\jpsi}(\pt) = \frac{N_{\jpsi}(\pt)}{\textrm{BR}_{\rb{ee}} \:\: N_{\rb{evts}}
    \:\: \langle\acceff\rangle(\pt)} \:.
\end{equation}
Here $\textrm{BR}_{\rb{ee}}$ is the branching ratio for the decay
$\jpsi \rightarrow \epem$, $N_{\rb{evts}}$ the number of events, and
$\langle\acceff\rangle$ the phase space dependent product of
acceptance $A$ and reconstruction efficiency $\epsilon$.  The latter
is calculated from Monte Carlo (MC) simulations as the ratio between
the number of reconstructed and generated MC \jpsi, which are assumed
to be unpolarized.  In pp collisions at \sqrts~= 7~TeV the \jpsi\
polarization has been measured and was found to be compatible with
zero at mid-rapidity ($\pt > 10$~\gevc) and forward rapidity ($\pt >
2$~\gevc) \cite{Abelev:2011md,Chatrchyan:2013cla,Aaij:2013nlm}.  In
heavy-ion collisions no measurement exists, but \jpsi\ mesons produced
from the recombination of charm quarks in the medium are expected to
be unpolarized\footnote{The impact of the polarization on the
  acceptance was studied for extreme polarization scenarios in
  \cite{Aamodt:2011gj}.}.
The MC events used for the calculation of $\langle\acceff\rangle$ are
constructed by adding to background events, generated with the HIJING
model \cite{Wang:1991hta}, \jpsi~mesons decaying into \epem~pairs,
whose phase space distribution is obtained from extrapolations of
other measurements \cite{Bossu:2011qe}, taking into account shadowing
effects as parameterized in EKS98 \cite{Eskola:1998df}.  The
dielectron decay is simulated with the EvtGen \cite{Lange:2001uf}
package, using the PHOTOS model \cite{Barberio:1993qi} to describe the
influence of final state radiation.  This choice, together with the
simulation of bremsstrahlung in the detector material, is mandatory
for a proper description of the low mass tail in the measured
\jpsi~mass distribution and ensures that the fraction of the signal
outside of the \minv\ integration window is properly accounted for in
the correction $\langle\acceff\rangle$.  The propagation of the
simulated particles is done by GEANT3 \cite{Brun:1994aa} and a full
simulation of the detector response is performed.  The same
reconstruction procedure and cuts are applied to MC events and to real
data.  The quality of the simulation is illustrated by the good
agreement of the background-subtracted invariant mass distributions
with the MC simulation of the \jpsi\ signal shape, after normalizing
it to the same integral as the measured signal (see
\Fi{fig:minvincl}).

The analysis has been performed in two slightly different centrality
intervals ($0$--$40$\% and $0$--$50$\%), where the larger one is used
for the extraction of non-prompt \jpsi\ which requires a higher
statistics than the inclusive measurement.  Also, the \pt~intervals
have been optimized for the different analyses.  It was checked that
the results for inclusive \jpsi\ obtained with the two centrality
binnings are in good agreement.

{\it Determination of the pp reference for \raa} \\
From the corrected \jpsi\ yield $Y_{\jpsi}(\pt)$ the nuclear
modification factor $\raa(\pt)$ is calculated as
\begin{equation}
\raa(\pt) = \frac{Y_{\jpsi}(\pt)}{\taa \:
  \sigma_{\jpsi}^{\rb{pp}}(\pt)} \:.
\end{equation}
Since no differential \jpsi\ measurement at mid-rapidity at low \pt\
is available for pp collisions at \sqrts~= 2.76~TeV
\cite{Abelev:2012kr}, the reference needed for the construction of
\raa\ is based on an interpolation of the mid-rapidity measurements by
PHENIX at \sqrts~= 0.2~TeV \cite{Adare:2006kf}, CDF at \sqrts~=
1.96~TeV \cite{Acosta:2004yw}, and ALICE at \sqrts~= 7~TeV
\cite{Abelev:2012kr}.  The interpolated \pt~distribution is obtained
by fitting the following parameterization to the available data sets
\cite{Bossu:2011qe}
\begin{equation}
\label{eq:bossu}
\frac{1}{\der \sigma/\der y} \frac{\der^{2} \sigma}{\der \zt \der y} =
c \: \frac{\zt}{(1 + a^{2} \zt^{2})^{n}} \:.
\end{equation}
Here, \zt\ is defined as $\pt / \ptavg$, $a = \Gamma(3/2)\; \Gamma(n -
3/2) / \Gamma(n - 1)$, and $c = 2 (n - 1)\; a$, where $n$ is the only
free fit parameter.  The value for \ptavg\ (calculated in the
\pt~range 0~--~10~\gevc) at \sqrts~= 2.76~TeV, which is needed to
translate this parameterization into $\der \sigma/\der \pt$, is
determined by interpolating between the existing \ptavg\ measurements
for pp and \ppbar\ collisions
\cite{Adare:2006kf,Acosta:2004yw,Abelev:2012kr}.  This interpolation
is done using various functional forms for the \sqrts~dependence to
determine the systematic uncertainty.  For the absolute normalization
of the parametrized spectrum, the same interpolated value $\der \sigma
/ \der y = 4.25 \pm 0.28\textrm{(stat.)} \pm
0.43\textrm{(syst.)}$~$\mu$b as in \cite{Abelev:2013ila} is used.

The main sources of systematic uncertainties for the \pt~dependent
\raa\ of inclusive \jpsi\ are the signal reconstruction procedure, the
MC input kinematics, the uncertainties on the interpolated pp
reference and on the nuclear overlap function.  The corresponding
values are summarized in \Ta{tab:systUncRaa}.  While the first two
components are uncorrelated between the \pt~intervals (type II), the
uncertainty due to the nuclear overlap function is fully correlated
(type I).  The pp reference on the other hand introduces both
uncorrelated and correlated contributions.  To determine the
uncertainty related to the signal reconstruction, the normalization
range of the ME background and the size and positions of the
\minv~bins have been varied.  All track and electron selection
criteria, such as the electron inclusion cut and the SPD hit
requirement, have been relaxed and/or tightened in order to test the
stability of the result, as was performed in \cite{Abelev:2013ila}.
The value of the systematic uncertainty is determined as the standard
deviation of the distribution of all results obtained with the listed
variations.  The evaluation of the uncertainties associated with the
MC input kinematics is also described in \cite{Abelev:2013ila}, while
the uncertainty of the pp reference is estimated from the differences
between the cross-section values obtained with the fitting procedure
based on \Eq{eq:bossu} and the measured values used for the fit at the
various energies.

%
\begin{figure}
\begin{center}
\includegraphics[width=0.95\linewidth]{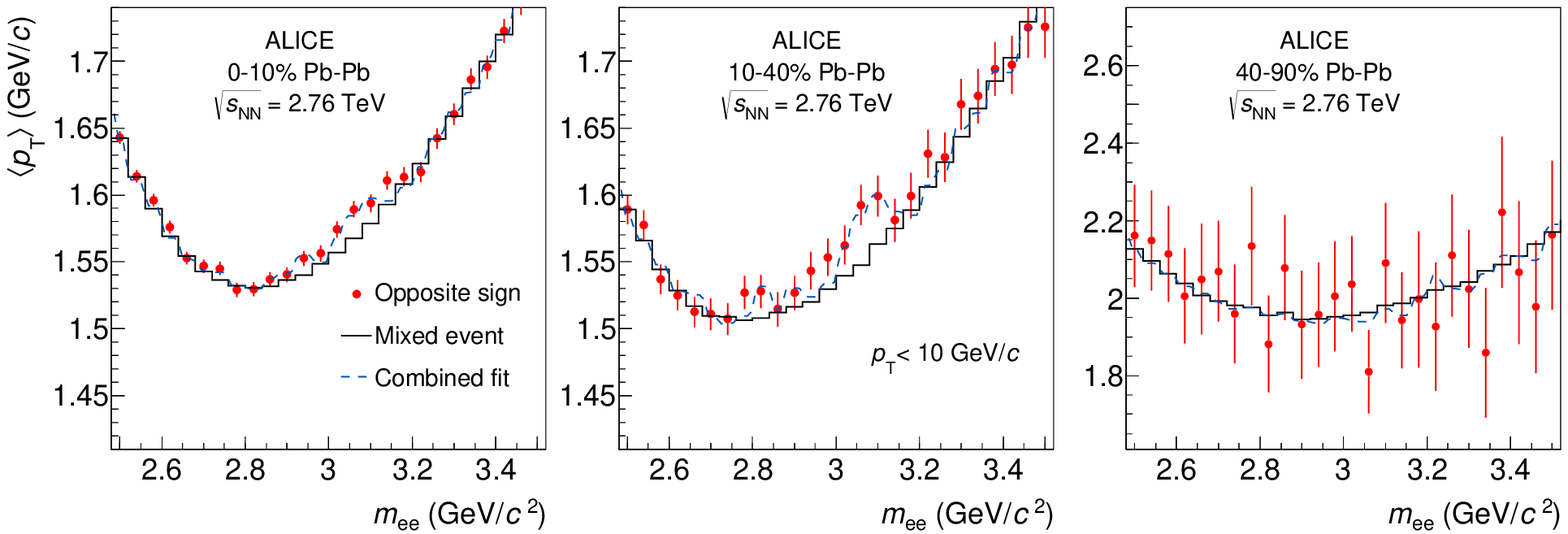}
\end{center}
\caption{\label{fig:meanptminv}
The average transverse momentum \ptavg\ of \epem~pairs, measured for
the \pt~range 0~--~10~\gevc, as a function of the invariant mass
\minv\ in centrality selected Pb-Pb collisions at \sqrtsnn~=
2.76~TeV.  The shown uncertainties are statistical only.  The
background \ptavg~distributions and the total fit results are also
shown superimposed to the data points.
}
\end{figure}
%

{\it Determination of \ptavg\ and \ptavgsq} \\
Since the collected Pb-Pb statistics would allow the extraction of the
\jpsi\ yield in a few \pt~intervals only, the average transverse
momentum \ptavg\ is determined by a fit to the distribution of the
\ptavg\ of \epem~pairs as a function of \minv.  When building such a
distribution, the individual \epem~pairs are weighted by the inverse
of their acceptance times efficiency $(\acceff)^{-1}$, assuming that
they come from the decay of a \jpsi.  The resulting \ptavg\
distributions are fitted by the expression
\begin{equation}
\label{eq:ptfit}
\ptavg_{\rb{meas}} = \frac{1}{S(\minv) + B(\minv)} \: \left[ S(\minv)
  \: \ptavg_{\jpsi} + B(\minv) \: \ptavg_{\rbt{Bkg}} \right] \:.
\end{equation}
Both factors $S$ and $B$ depend on \minv\ and correspond to the
distribution of the \jpsi~signal and of the background.  For $S$ the
same background subtracted signal distribution $S(\minv)$ is used as
for the extraction of the yield (see lower panels of
\Fi{fig:minvincl}), while the background $B$ is generated from the ME
sample, as $B(\minv) = c_{\rbt{B}} \: B_{\rbt{ME}}(\minv)$.  The
normalization factor is determined by fitting $c_{\rbt{B}} \:
B_{\rbt{ME}}(\minv)$ to the corresponding \minv~distribution of
\epem~pairs in the regions $1.5 < \minv < 2.5$~\gevcc\ and $3.2 <
\minv < 4.2$~\gevcc, thus excluding the signal region.  For the sum
$S(\minv) + B(\minv)$ in the denominator of \Eq{eq:ptfit}, the
measured OS pair \minv~distribution is used.  The $\ptavg_{\rbt{Bkg}}$, 
defined as the \ptavg\ of the combinatorial background pairs, is also
calculated from the ME sample.  This analysis is performed in three
different centrality intervals: $0$--$10$\%, $10$--$40$\%, and
$40$--$90$\%.  Figure~\ref{fig:meanptminv} shows the measured \ptavg\
of the \epem~pairs in the \pt~range 0~--~10~\gevc\ together with the
results of the fit procedure.  In addition, with an equivalent method,
the mean square transverse momentum \ptavgsq\ is also calculated for
the same centrality intervals.

The systematic uncertainties of the \ptavg\ measurement for inclusive
\jpsi\ are mainly determined by the signal extraction, the stability
of track and electron selection criteria and the fit procedure (see
\Ta{tab:systUncPt}).  While the first two components are not
correlated between the different centrality intervals (type II), the
systematic uncertainty intrinsic to the fit procedure can affect the
data points in a correlated way (type I).  The uncertainties related
to the signal extraction have been evaluated by varying the
normalization range of the ME background, the size and positions of
the \minv~bins, the fit region and by using in addition to the ME
background a linear function for the background description.  In
addition, the stability of the fit procedure was tested by modifying
the approach, e.g. by using in the fit the direct sum of $S$ and $B$,
instead of the OS pair \minv~distribution, or by using a fit function
for $B$, instead of ME.  It was also verified by applying the above
described method to MC events, which were constructed by combining
signal with background events with a realistic $S/B$~ratio.  It turned
out that the procedure allows to recover the \ptavg\ of the simulated
\jpsi~mesons within a 2\% difference.  This value is assumed as
correlated (type I) uncertainty.  Finally, the uncertainty in the
signal-to-background ratio is propagated into the statistical
uncertainty of the $\ptavg_{\jpsi}$.

%
\begin{table}[t]
\centering
\begin{tabular}{|l|c|c|c|c|c|c|}
                                                                             \hline
                               & \multicolumn{5}{c|}{ \pt~range (\gevc) } & \\
                               & \multicolumn{2}{c|}{ $0$--$40$\% }
                               & \multicolumn{3}{c|}{ $0$--$50$\% }         &  \\ 
Source                         & 0~--~2.5 & 2.5~--~6 & 0~--~1.5 & 1.5~--~4.5 & 4.5~--~10 & Type \\ \hline
Signal reconstruction          & 10.9     & 11.0     & 22.6     &  8.5       & 10.0      & II   \\
MC input kinematics            &  3.2     &  5.9     &  1.5     &  4.7       &  5.3      & II   \\
Nuclear overlap function \taa\ &  3.2     &  3.2     &  3.8     &  3.8       &  3.8      & I    \\ \hline
\multirow{2}{*}{pp reference}  & 12.1     & 12.1     & 12.1     & 12.1       & 12.1      & I    \\
                               &  4.5     &  4.9     &  5.0     &  4.3       & 10.1      & II   \\ \hline
\multirow{2}{*}{Total}         & 12.6     & 12.6     & 12.7     & 12.7       & 12.7      & I    \\
                               & 12.3     & 13.5     & 23.2     & 10.6       & 15.2      & II   \\ \hline
\end{tabular}
\caption{
\label{tab:systUncRaa}
The correlated (type I) and uncorrelated (type II) systematic
uncertainties (in percent) on the measurement of the nuclear
modification factor \raa\ of inclusive \jpsi\ for several
\pt~intervals in Pb-Pb collisions ($0$--$40$\% and $0$--$50$\% most
central) at \sqrtsnn~= 2.76~TeV.
}
\end{table}
%

%
\begin{table}[t]
\centering
\begin{tabular}{|l|c|c|c|c|}
                                                                  \hline
                                         & \multicolumn{3}{c|}{ Centrality }  &      \\
Source                                   & $0$--$10$\% & $10$--$40$\% & $40$--$90$\% & type \\ \hline
Signal extraction                        &      2.0  &      1.6   &      4.5   & II   \\
Track selection                          &      3.1  &      3.9   &      13.7  & II   \\
Stability of the fit procedure (from MC) &      2.0  &      2.0   &      2.0   & I    \\ \hline
\multirow{2}{*}
{Total}                                  &      2.0  &      2.0   &      2.0   & I    \\
                                         &      3.7  &      4.2   &      14.4  & II   \\ \hline
\end{tabular}
\caption{
\label{tab:systUncPt}
The correlated (type I) and uncorrelated (type II) systematic
uncertainties (in percent) on the measurement of the average
transverse momentum \ptavg\ of inclusive \jpsi\ in three centrality
intervals in Pb-Pb collisions at \sqrtsnn~= 2.76~TeV.
}
\end{table}
%

%
\begin{figure}
\begin{center}
\begin{minipage}[b]{0.60\linewidth}
\includegraphics[width=0.96\linewidth]{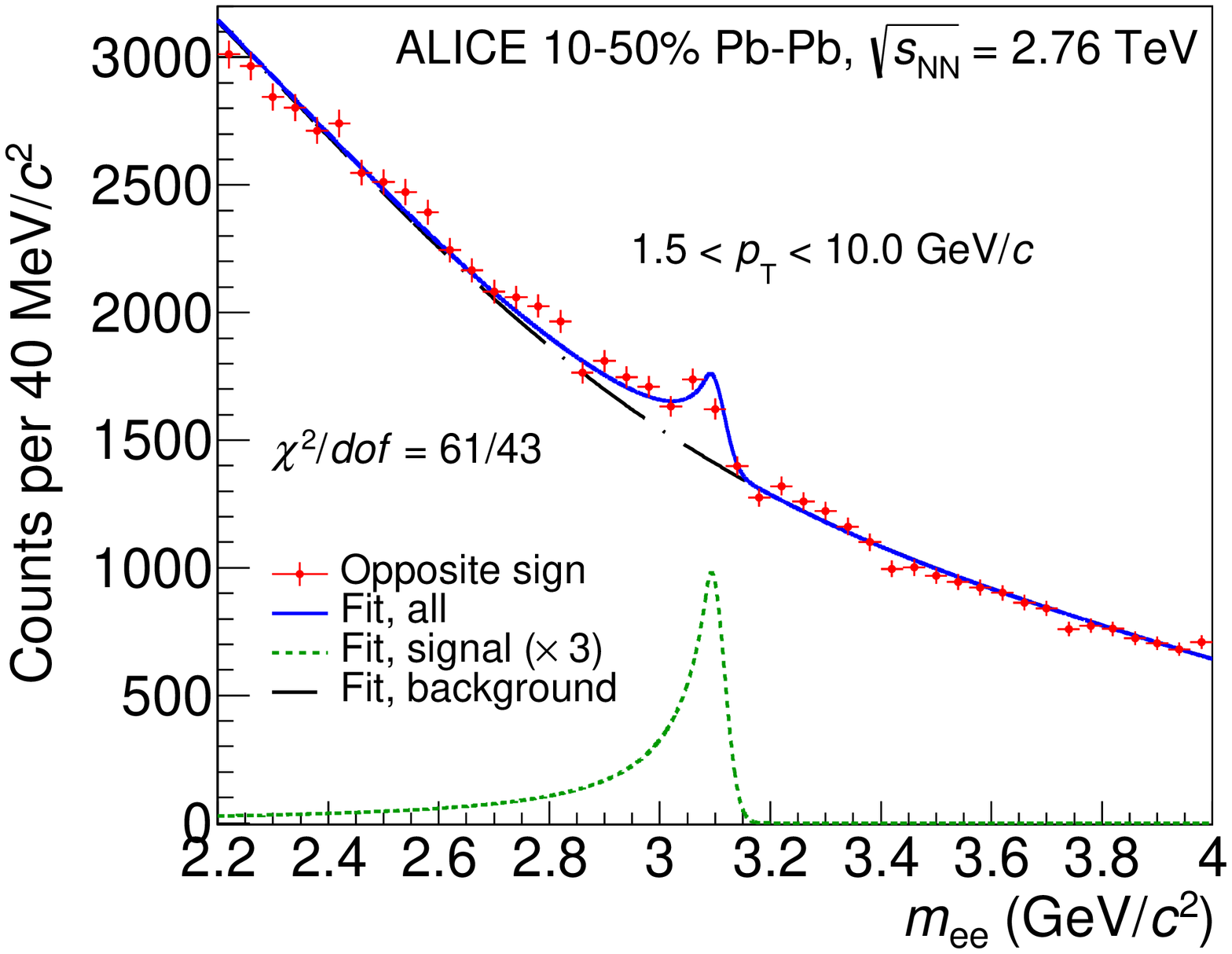}
\end{minipage}
\begin{minipage}[b]{0.60\linewidth}
\includegraphics[width=1.0\linewidth]{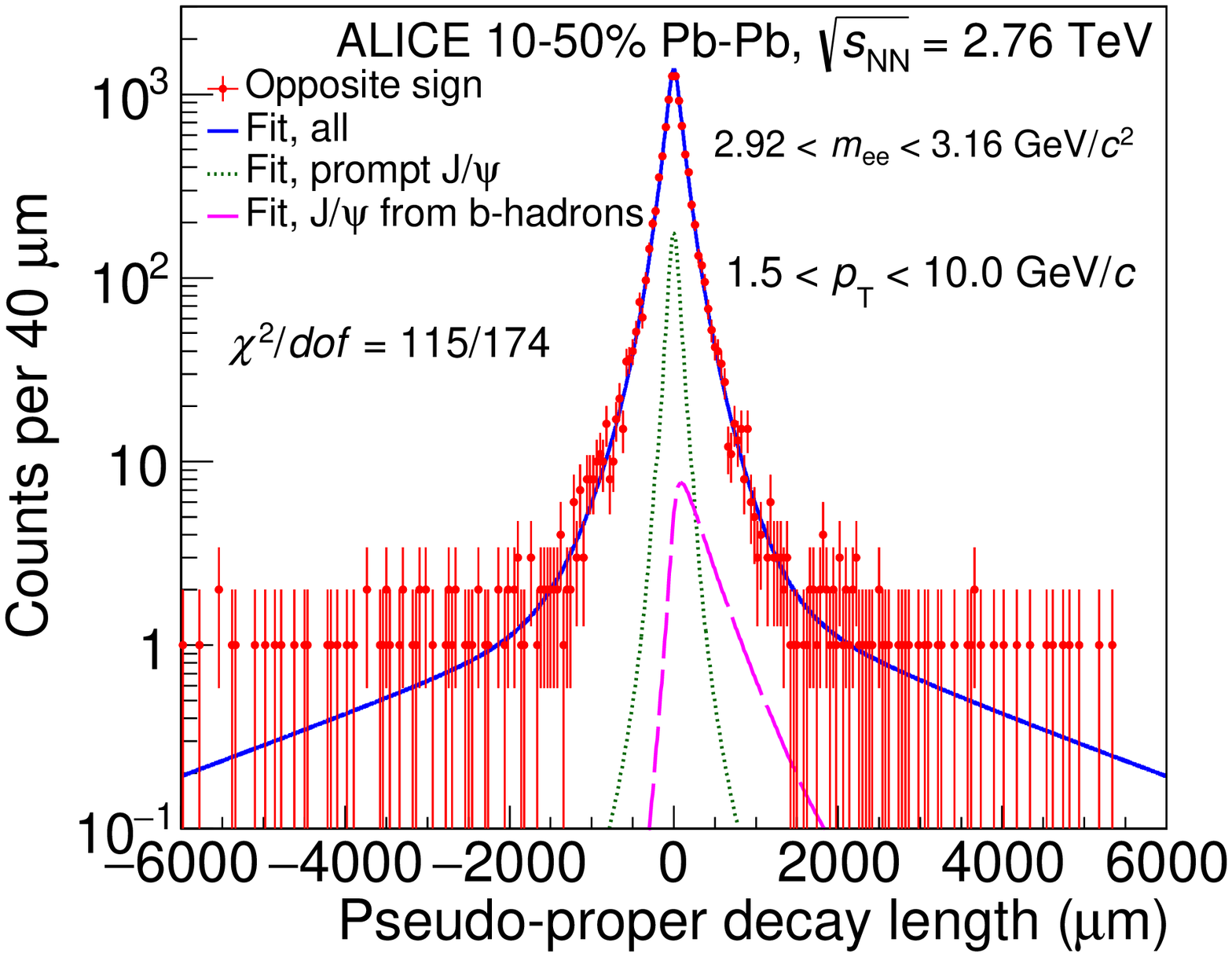}
\end{minipage}
\end{center}
\caption{\label{fig:nonpromptfit}
The invariant mass (upper panel) and pseudo-proper decay length (lower
panel) distributions for \epem\ pairs with $\pt > 1.5$~\gevc\ in Pb-Pb
collisions in the centrality interval $10$--$50$\% at \sqrtsnn~=
2.76~TeV.  The projections of the maximum likelihood fit used to
extract \fb\ are superimposed to the data.}
\end{figure}
%

%
\subsection{Non-prompt \jpsi}

The candidate selection for the non-prompt \jpsi\ analysis includes,
in addition to the previously described criteria, the condition that
at least one of the two decay tracks has a hit in the first SPD layer,
in order to enhance the resolution of secondary vertices.

The non-prompt \jpsi~fraction has been determined using an unbinned
two-dimensional log-likelihood fit described in detail in
\cite{Abelev:2012gx}, which is performed by maximizing the quantity  
\begin{equation}
\label{logLikeFunc}
  \ln L = \sum^{N}\ln \left[ \fsig 
             \cdot \ffs(x) 
             \cdot M_{\rbt{Sig}}(\minv) + (1 - \fsig) 
             \cdot \ffb(x) 
             \cdot M_{\rbt{Bkg}}(\minv) \right], 
\end{equation}
where $N$ is the total number of OS candidates in the range $2.2 <
\minv < 4$~\gevcc\ and $x$ is the pseudo-proper decay length of the
candidate
\begin{equation}
   x = \frac{c \: (\vec{L} \cdot \vec{\pt}/\pt) \: m_{\jpsi}}{\pt}. 
\end{equation}
Here $\vec{L}$ is the vector pointing from the primary vertex to the
\jpsi~decay vertex and $m_{\jpsi}$ the mass of the \jpsi\ taken from
\cite{Agashe:2014kda}.  $\ffs(x)$ and $\ffb(x)$
($M_{\rbt{Sig}}(\minv)$ and $M_{\rbt{Bkg}}(\minv)$) are Probability
Density Functions (p.d.f.) describing the pseudo-proper decay length
(invariant mass) distribution for signal and background candidates,
respectively.  $\ffs(x)$ is defined as
\begin{equation}
  \ffs(x) = \fbprime \cdot F_{\rbt{B}}(x) + (1 - \fbprime) 
                     \cdot F_{\rb{prompt}}(x) \: ,
\end{equation}
where $F_{\rb{prompt}}(x)$ and $F_{\rbt{B}}(x)$ are the p.d.f. for prompt 
and non-prompt \jpsi, respectively, and \fbprime\ is the fraction of
{\it reconstructed} \jpsi\ coming from beauty hadron decays
\begin{equation}
  \fbprime = \frac{N_{\jpsi \leftarrow h_{B}}}
                  {N_{\jpsi \leftarrow h_{B}} 
                 + N_{\rb{prompt}}} \:\:\:\:.
\end{equation}
A correction due to different average $\langle \acceff \rangle$
values, in a given \pt\ interval, for prompt and non-prompt \jpsi, is
necessary to obtain from \fbprime\ the fraction of {\it produced}
non-prompt \jpsi, \fb
\begin{equation}
\label{eq:fbCorr}
  \fb = \left (1 + \frac{1 - \fbprime}{\fbprime} 
        \frac{\langle \acceff \rangle_{\rbt{B}}}
             {\langle \acceff \rangle_{\rb{prompt}}} \right )^{-1} \:.
\end{equation}

The various ingredients for the determination of \fb\ are described in
the following:

{\it Monte Carlo \pt\ distributions and polarization assumptions} \\ 
Assuming both prompt and non-prompt \jpsi\ to be unpolarized, at a
given \pt\ their acceptance times efficiency values
$\langle\acceff\rangle$ are the same.  However, the \pt\ distributions
of prompt and non-prompt \jpsi\ can be different, resulting in
different average $\langle\acceff\rangle$ computed over a \pt\ range
of finite size.  Different hypotheses for the kinematical (\pt)
distributions of both prompt and non-prompt \jpsi\ are considered,
i.e. including or excluding shadowing or suppression effects as, e.g.,
those predicted in references
\cite{Alberico:2011zy,Alberico:2013bza,Uphoff:2012gb} for non-prompt
\jpsi.  Due to the weak \pt\ dependence of $\langle\acceff\rangle$,
the resulting uncertainty on \fb\ is small, being $\sim 5$\% at low
\pt\ and $\sim 3$\% in the highest \pt\ bin, and is independent of
centrality.

At a given \pt, prompt and non-prompt \jpsi\ can have different
polarization and therefore a different acceptance.  However, the
polarization of \jpsi\ from b-hadron decays is expected to be small
due to the averaging effect caused by the admixture of various
exclusive B~$\rightarrow \jpsi + X$ decay channels.  Indeed, in more
elementary colliding systems, the sizable polarization, which is
observed when the polarization axis refers to the B-meson direction
\cite{Aubert:2002hc}, is strongly smeared when calculated with respect
to the direction of the daughter \jpsi\ \cite{Aaij:2011jh}, as
observed by CDF \cite{Abulencia:2007us}.  The central values of the
fraction of non-prompt \jpsi\ are evaluated with \Eq{eq:fbCorr}
assuming unpolarized prompt \jpsi\ and a polarization of non-prompt
\jpsi\ as predicted by EVTGEN \cite{Lange:2001uf}.  The assumption of
a null polarization for non-prompt \jpsi\ results in a relative
decrease of \fb\ by only 1\% at high \pt\ (4.5~--~10~\gevc) and 3\%
at low \pt\ (1.5~--~4.5~\gevc).  The relative variations of \fb\
expected in extreme scenarios for the polarization of prompt \jpsi\ was
studied in \cite{Aamodt:2011gj}.  The uncertainties related to the
polarization of prompt and non-prompt \jpsi\ are are not further
propagated to the results.

{\it P.d.f. for prompt \jpsi: $F_{\rb{prompt}}(x)$} \\
The $x$ distribution $F_{\rb{prompt}}(x)$ for prompt \jpsi, which
decay at the primary vertex, coincides with the resolution
function $R(x)$, which describes the accuracy by which $x$ can be
reconstructed.  It also enters in the p.d.f. describing the $x$
distributions of non-prompt \jpsi, $F_{\rbt{B}}(x)$, and of the
background candidates, $\ffb(x)$.  The determination of $R(x)$ is
based on the same MC data sample as used for the inclusive \jpsi\
analysis (see Sect.~\ref{sec:incljpsirec}).  The systematic
uncertainty on $R(x)$ was estimated with a MC approach by propagating
the maximum observed discrepancies of the track parameters (space and
momentum variables) between data and MC to the $x$ variable 
\cite{Abelev:2014ffa,ALICE:2012ab,Abelevetal:2014dna} and was found to
be at most 10\%.  To propagate this systematic uncertainty to the
final results the fits are repeated after modifying in the
log-likelihood function the resolution to $\left( 1 / (1+\delta)
\right) \times R \left( x / (1+\delta) \right)$.  In this expression
$\delta$ parameterizes the relative variation of the RMS of the
resolution function and is varied between $-10$\% and $+10$\%.  The
systematic uncertainty due to the resolution function is smaller in
the highest \pt\ bin, because of the better resolution in the $x$
variable and the higher values of the signal-to-background ratio. 

{\it P.d.f. for non-prompt \jpsi: $F_{\rbt{B}}(x)$} \\  
The shape of the $x$ distribution of non-prompt \jpsi\ is estimated by
using PYTHIA 6.4.21 \cite{Sjostrand:2006za} in the Perugia-0 tune
\cite{Skands:2010ak} to generate beauty hadrons at \sqrts~= 2.76~TeV,
and the EvtGen package~\cite{Lange:2001uf} to describe their decays.
The systematic uncertainty related to this shape is estimated by
assuming a softer \pt~distribution for the non-prompt \jpsi\, which is
obtained by adding the suppression effects as predicted
in~\cite{Uphoff:2012gb} and a harder one taken from the same PYTHIA
event generator at \sqrts~= 7~TeV instead of 2.76~TeV.  The resulting
systematic uncertainty is within 3~--~4\%.

{\it P.d.f. for the background: $\ffb(x)$} \\
The main difference of the analysis presented in this section, with
respect to previous work on pp collisions~\cite{Abelev:2012gx},
concerns the description of $\ffb(x)$.  In this analysis such a
function includes an extra symmetric exponential tail ($\propto
e^{-|x|/\lambda_{\rm sym}} $) \cite{Acosta:2004yw} and depends on the
invariant mass and the \pt\ of the dielectron pair.  It is determined,
for each centrality class, by a fit to the data in three \pt\ regions
(1.5~--~3, 3~--~4.5, 4.5~--~10~\gevc) and in four invariant mass
regions on the side-bands of the \jpsi\ mass peak (2.2~--~2.6,
2.6~--~2.8, 3.16~--~3.5, 3.5~--~4~\gevcc, labelled with the indices 1,
2, 3 and 4, respectively), for a total of $3 \times 4$ combinations.
The background function in the invariant mass region
2.8~--~3.16~\gevcc\ and in each of the three \pt\ ranges are obtained
by an interpolation procedure as the weighted combination of the
p.d.f. determined in the other four invariant mass regions.  The
weights are chosen inversely proportional to the absolute difference
(or its square) between the mean of the invariant mass distribution in
the given mass interval and that in the interpolated region
\begin{equation}
{F_{{\rm Bkg}}}_{\rm interp}  (x) = 
      \sum_{i=1}^{4} w_{i} \: {F_{{\rm Bkg}}}_{i} (x) ; \quad 
       w_{i} \propto | \langle \minv \rangle_{i} - 
                      \langle \minv \rangle_{\rm interp} |^{-n} 
\; \; \; \; \; \; (n \; = \; 1 \; {\rm  or} \; 2).
\label{weights}
\end{equation}
Optionally, only the two adjacent mass regions can be considered in the 
interpolation procedure, corresponding to the condition  $w_{1} =
w_{4} = 0$.  The central value of \fb\ has been determined as the
average of the values obtained with the different assumptions ($n = 1$
or $n = 2$, with or without the condition $w_{1} = w_{4} = 0$).  The
RMS of the distributions of the relative variations obtained for \fb\
is used to define the systematic uncertainty.  It becomes larger for
central events and in the lowest \pt\ interval, where the
signal-to-background ratio $S/B$ is lower.  This approach allows to
cope with the much lower $S/B$ ratio in Pb-Pb than in pp collisions.

{\it P.d.f. for the invariant mass distribution of the signal:
  $M_{\rbt{Sig}}(\minv)$} \\
The shape of the invariant mass distribution for the signal is
determined by the same MC simulations described in
Sect.~\ref{sec:incljpsirec}.  The influence of detector material
budget is studied with dedicated MC simulations, where the material
budget is varied within its uncertainty ($\pm 6$\%)
\cite{Koch:2011fw}.  The resulting contribution to the systematic
uncertainty on \fb\ slightly increases for central events, and ranges
from 2 to 4\%. 

{\it P.d.f. for the invariant mass distribution of the background:
  $M_{\rbt{Bkg}}(\minv)$} \\ 
The shape of the invariant mass distribution for the background
candidates is determined from ME pairs.  The related systematic
uncertainty on \fb\ is evaluated using the like-sign distribution,
instead of the ME one.  The uncertainty increases at higher centrality
and in the lowest \pt\ interval due to the decrease of the S/B ratio. 

As an example in \Fi{fig:nonpromptfit} the projections of
the best fit function for $n = 1$\ and $w_{1} = w_{4} = 0$\ are shown
superimposed to the invariant mass (upper panel) and $x$ (lower panel)
distributions of the candidates in the centrality range $10$--$50$\%
for $1.5 < \pt < 10$~\gevc.  

A summary of the systematic uncertainties on the determination of the
non-prompt \jpsi\ fraction is provided in \Ta{tab:systUncNonP} for the
three centrality intervals in the integrated \pt\ range and, in the
two \pt\ ranges where the results will be given, for the most central
collisions ($0$--$10$\%).

The value of \fb\ is determined in two \pt\ bins (1.5~--~4.5 and
4.5~--~10~\gevc) for the $0$--$50$\% centrality range and in three
centrality classes ($0$--$10$\%, $10$--$40$\% and $40$--$90$\%) for
$1.5 < \pt < 10$~\gevc.  The \fb\ measurements are then combined with
the nuclear modification factors of inclusive \jpsi\ to get the
non-prompt and prompt \jpsi\ \raa
\begin{equation}
\label{eq:raa-non-prompt}
  \raa^{\rm non-prompt\ \jpsi} = \frac{\fb^{\rm Pb-Pb}}{\fb^{\rm pp}}
                         \ \raa^{\rm incl.\ \jpsi} 
                         \ \ , \ \ \ \ \ \ \ \ \ \ \ \ \ \
  \raa^{\rm prompt\ \jpsi} = \frac{1 - \fb^{\rm Pb-Pb}}{1 - \fb^{\rm pp}}
                       \ \raa^{\rm incl.\ \jpsi} \:.
\end{equation}

{\it pp interpolation} \\
The value of \fb\ in pp collision at \sqrts~= 2.76~TeV, $\fb^{\rm
  pp}$, is needed to compute the \raa\ for prompt and non-prompt
\jpsi~mesons, see \Eq{eq:raa-non-prompt}.  It is determined by
an interpolation procedure.  Therefore, a fit is performed to the
existing measurements of \fb\ as a function of \pt\ in mid-rapidity pp
collisions at \sqrts~= 7~TeV (ALICE~\cite{Abelev:2012gx},
ATLAS~\cite{Aad:2011sp} and CMS~\cite{Khachatryan:2010yr}).  The
function used to fit the data is chosen as 
\begin{equation}
\label{fbmodel}
  \fb^{\rb{model}}(\pt) = \frac{d\sigma^{\rbt{FONLL}}_{\jpsi
                              \leftarrow h_{\rm B}}}{d\pt}
                        \left/
                        \frac{d\sigma_{\jpsi}^{\rb{phenom.}}}
                             {d\pt}
                        \right., 
\end{equation}
which is the ratio of the differential cross section for non-prompt
\jpsi\ obtained by an implementation of pQCD calculations at fixed
order with next-to leading-log resummation (FONLL)
\cite{Cacciari:2012ny} to that for inclusive \jpsi, parameterized by
the phenomenological function defined in \Eq{eq:bossu}.  A similar fit
is then performed to the CDF results \cite{Acosta:2004yw} in \ppbar\
collisions at \sqrts~= 1.96~TeV.  Finally, the $\fb^{\rm pp}$(\pt)
value at \sqrts~= 2.76~TeV is determined by an energy interpolation,
which gives $\fb^{\rm pp} = 0.122 \pm 0.010$ in the integrated \pt\
range 1.5~--~10 \gevc.  The quoted uncertainty includes: {\it(i)} a
component from the fit procedure, which depends on the uncertainties
of both data and FONLL predictions;
{\it (ii)} the systematic uncertainty due to the energy interpolation,
which has been estimated by considering different functional forms of
the \sqrts\ dependency (linear, exponential and power law); 
{\it (iii)} an additional systematic uncertainty, which has been
obtained by repeating the whole fitting procedure after excluding, one
at a time, the data samples used for the \fb\ fit in pp collisions at
\sqrts~= 7~TeV.

%
\begin{figure}
\begin{center}
\includegraphics[width=0.9\linewidth]{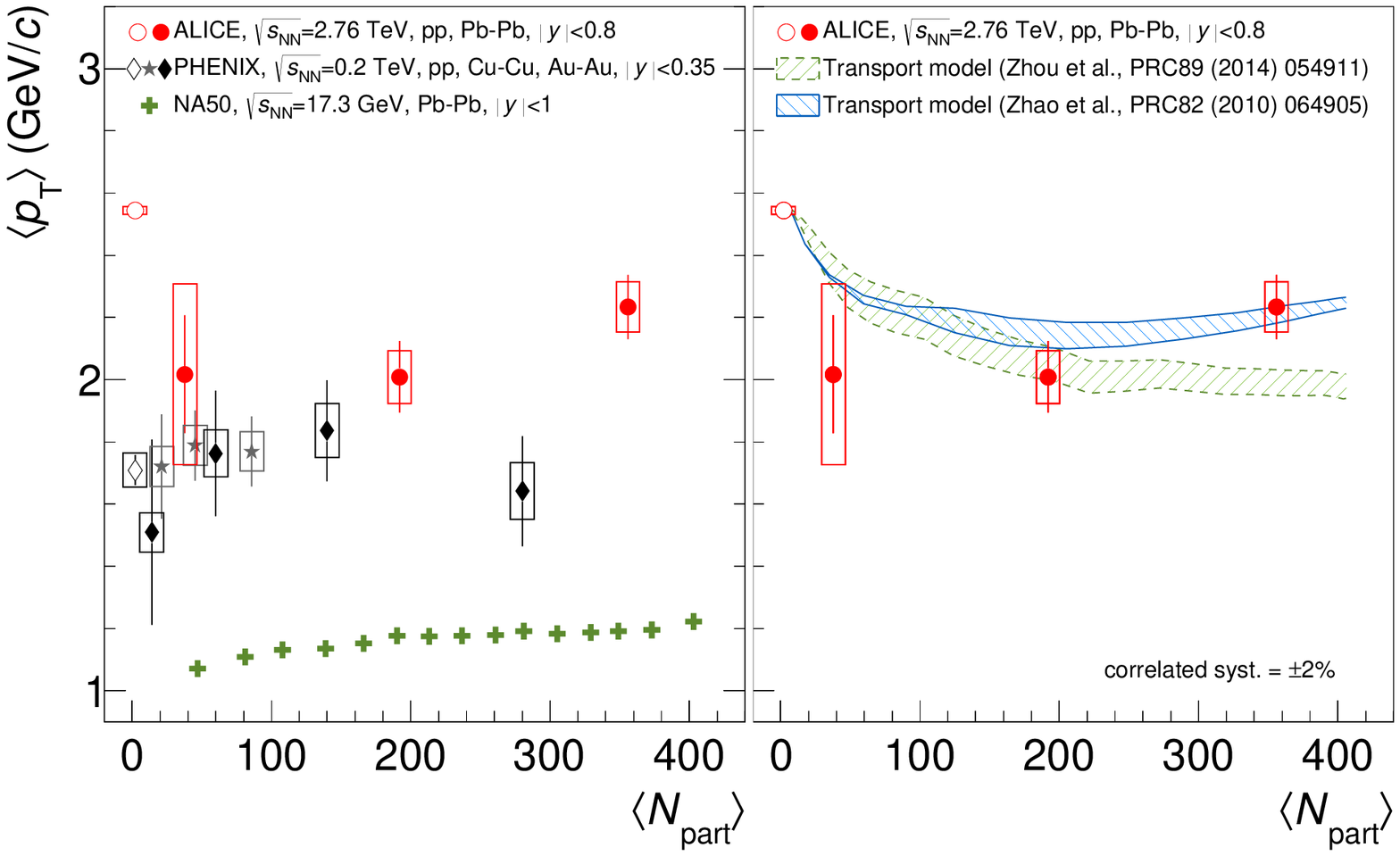}
\end{center}
\caption{
The average transverse momentum \ptavg\ of inclusive \jpsi\ measured
at mid-rapidity ($|y| < 0.8$) in centrality selected Pb-Pb collisions
(filled circles) and pp collisions (open circles) at \sqrtsnn~=
2.76~TeV as a function of the number of participants \npart.  The
uncorrelated systematic uncertainties (type II) are depicted by the
open boxes.  
Left panel: A comparison to results obtained by the PHENIX
collaboration for Au-Au and Cu-Cu collisions at \sqrtsnn~= 0.2~TeV
\cite{Adare:2006ns,Adare:2008sh} (open and filled diamonds) and by the
NA50 collaboration for Pb-Pb collisions at \sqrtsnn~= 17.3~GeV
\cite{Abreu:2000xe} (crosses).  The \ptavg\ values are calculated for
NA50 and PHENIX in the \pt~interval 0~--~5~\gevc, while for ALICE the
\pt~interval is 0~--~10~\gevc.
Right panel: \ptavg\ is compared to theory predictions by Zhou et
al. \cite{Zhou:2014kka} and Zhao et al. \cite{Zhao:2010nk,Zhao:2011cv}
for the \pt~interval 0~--~10~\gevc.}
\label{fig:syssizemeanpt}
\end{figure}
%

%
\begin{table}[t]
\centering
\resizebox{\columnwidth}{!}{
\begin{tabular}{|l|c|c|c|c|c|c|}
\hline
  & \multicolumn{3}{c|}{$1.5 < \pt < 10$~\gevc} 
  & \multicolumn{2}{c|}{Centr. $0$--$10$\%}     & Type \\
Source & Centr. & Centr. & Centr. & \pt & \pt   &      \\
  & $0$--$10$\% & $10$--$40$\% & $40$--$90$\% 
  & 1.5~--~4.5~\gevc & 4.5~--~10~\gevc          &      \\ \hline
Resolution function                         & 23 & 15 & 10 & 28 & 12 &  I \\
P.d.f. for non-prompt \jpsi                 &  4 &  3 &  3 &  4 &  3 &  I \\
P.d.f. for the background                   & 22 & 15 &  5 & 23 & 19 & II \\
MC \pt\ distribution                        &  5 &  5 &  5 &  5 &  3 &  I \\
P.d.f. for the invariant mass of signal     &  5 &  3 &  2 &  5 &  3 &  I \\
P.d.f. for the invariant mass of background &  7 &  5 &  3 &  7 &  5 & II \\
\hline
Total                                       & 34 & 23 & 13 & 38 & 24 &    \\
\hline
\end{tabular}
}
\caption{
Systematic uncertainties (in percent) on the measurement of the
fraction \fb\ of \jpsi\ from the decay of beauty hadrons, for
different centrality intervals in the transverse momentum range $1.5 <
\pt < 10$~\gevc, and in the two \pt\ intervals for the most central
collisions.  The contributions which are fully correlated between the
different centrality classes are denoted as type I, the uncorrelated
ones as type II.
\label{tab:systUncNonP}
}
\end{table}
%

%
\section{Results}

%
\begin{table}[t]
\centering
\begin{tabular}{|r|c|c|}
  \hline
  Centrality  & \ptavg\ (\gevc) & \ptavgsq\ (\gevcsq) \\
  \hline
   $0$--$10$\% & $2.23 \pm 0.10 \pm 0.08$ 
             & $5.50 \pm 0.58 \pm 0.25$ \\
  $10$--$40$\% & $2.01 \pm 0.12 \pm 0.08$
             & $4.97 \pm 0.65 \pm 0.34$ \\
  $40$--$90$\% & $2.02 \pm 0.19 \pm 0.29$
             & $5.15 \pm 1.05 \pm 1.23$ \\
  \hline
  pp         & $2.54 \pm 0.02 \pm 0.01$
             & $9.07 \pm 0.15 \pm 0.07$ \\
  \hline
\end{tabular}
\caption{The numerical values of \ptavg\ and \ptavgsq\ calculated in
  the range $0 < \pt < 10$~\gevc\ for the three analyzed centrality
  intervals in Pb-Pb collisions (the first uncertainty is the
  statistical and the second is the uncorrelated systematic (type II),
  the correlated uncertainty has a value of 2\%, see
  \Ta{tab:systUncPt}).  The values for pp collisions obtained by the
  interpolation procedure are given as a reference.}
\label{tab:ptavg}
\end{table}
%

Figure~\ref{fig:syssizemeanpt} shows the \ptavg\ of inclusive \jpsi\
for the three analyzed centrality intervals.  The numerical values for
\ptavg\ are summarized in \Ta{tab:ptavg}.  As a reference, the \ptavg\
in pp collisions at the same centre-of-mass energy, as determined by
the interpolation method described in Sect.~\ref{sec:incljpsirec}, is
also presented.  The \ptavg\ for Pb-Pb collisions is significantly
smaller than that for pp collisions.  Such a behaviour is not observed
at smaller centre-of-mass energies (see left panel of
\Fi{fig:syssizemeanpt}), for which no significant system size
dependence of \ptavg\ is seen.  This might indicate the onset of
processes which either deplete the high \pt\ region or enhance the
\jpsi\ production at low \pt\ in heavy-ion collisions at the LHC.  The
latter effect would be expected as a consequence of a significant
contribution from \ccbar\ coalescence.

%
\begin{figure}
\begin{center}
\includegraphics[width=0.9\linewidth]{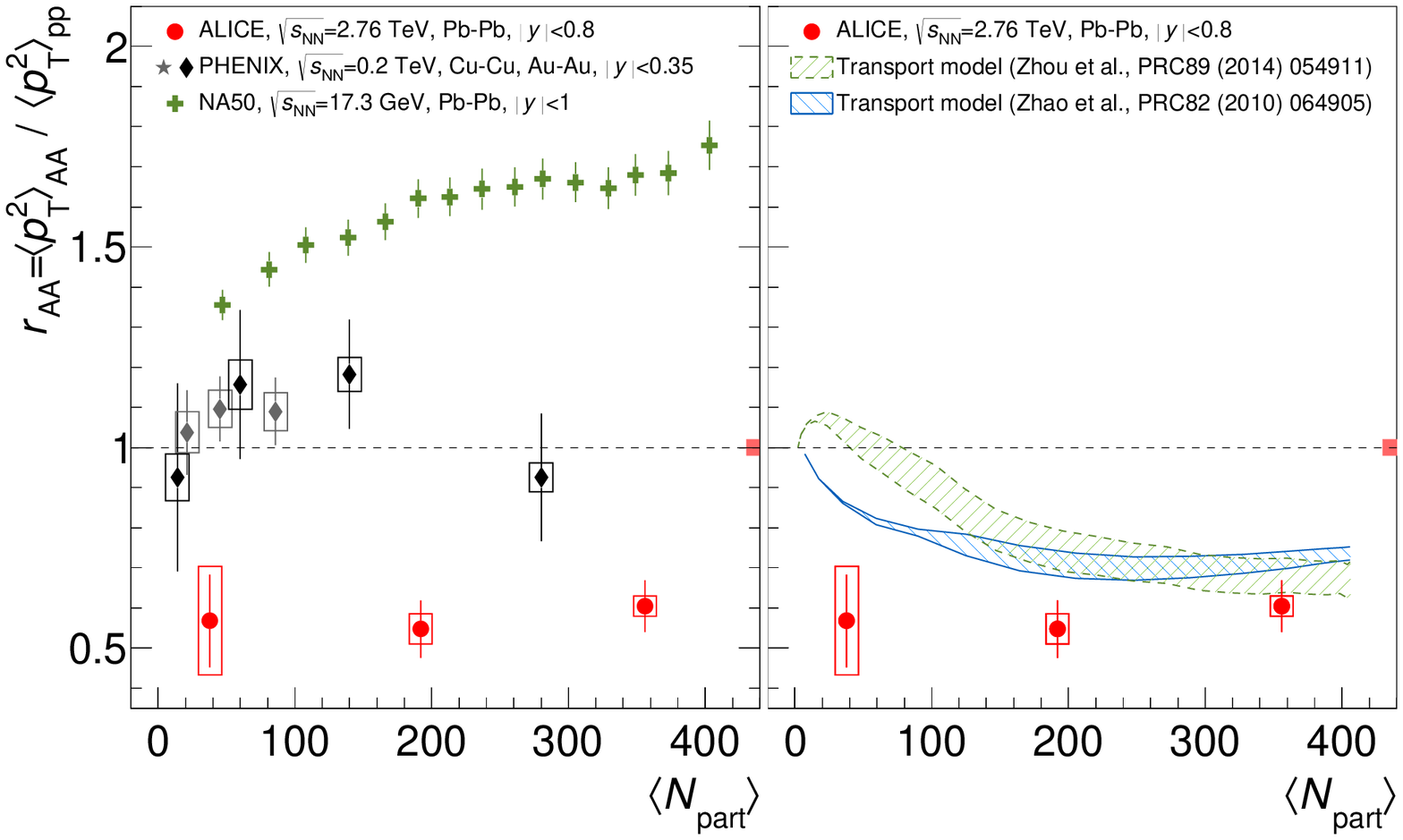}
\end{center}
\caption{
The ratio $\ptsqraa = \ptavgsq_{\rbt{AA}} / \ptavgsq_{\rbt{pp}}$ in
the \pt~interval 0~--~10~\gevc\ for inclusive \jpsi\ measured at
mid-rapidity ($|y| < 0.8$) in centrality selected Pb-Pb collisions
(filled circles) at \sqrtsnn~= 2.76~TeV as a function of the number of
participants \npart.  The uncorrelated systematic uncertainties (type
II) are depicted by the open boxes, while correlated uncertainty (type
I) is shown as the filled box at unity.
Left panel: A comparison to results obtained by the PHENIX
collaboration for Au-Au and Cu-Cu collisions at \sqrtsnn~= 0.2~TeV
\cite{Adare:2006ns,Adare:2008sh} (filled diamonds) and by the NA50
collaboration for Pb-Pb collisions at \sqrtsnn~= 17.3~GeV
\cite{Abreu:2000xe} (crosses).  The PHENIX and NA50 \ptsqraa\ values
are calculated in the \pt~interval 0~--~5~\gevc.
Right panel: \ptsqraa\ is compared to theory predictions by Zhou et
al. \cite{Zhou:2014kka} and Zhao et al. \cite{Zhao:2010nk,Zhao:2011cv}
for the \pt~interval 0~--~10~\gevc.}
\label{fig:raameanpt}
\end{figure}
%

It has been suggested \cite{Zhou:2013aea} that the observable
$\ptsqraa = \ptavgsq_{\rbt{AA}} / \ptavgsq_{\rbt{pp}}$ should be
particularly sensitive to medium modifications affecting the \jpsi\
transverse momentum distributions.  The measured \ptavgsq\ values for
Pb-Pb collisions at \sqrtsnn~= 2.76~TeV are summarized in
\Ta{tab:ptavg}.  The corresponding \ptsqraa\ values as a function of
\npart\ are shown in \Fi{fig:raameanpt} and are found to be
significantly below unity.  This is in contrast to results from lower
centre-of-mass energies, where either values consistent with unity
(PHENIX at \sqrtsnn~= 0.2~TeV \cite{Adare:2006ns,Adare:2008sh}) or
around 1.5 (NA50 at \sqrtsnn~= 17.3~GeV \cite{Abreu:2000xe}) were
obtained (see left panel of \Fi{fig:raameanpt}).  The measured \npart\
dependences of \ptavg\ and \ptsqraa\ are compared with a transport
model for inclusive \jpsi\ by Zhao et
al. \cite{Zhao:2010nk,Zhao:2011cv} in the right panels of
\Fis{fig:syssizemeanpt}{fig:raameanpt}.  This model includes
regeneration and dissociation processes, based on in-medium \jpsi\
spectral functions, throughout the evolution of a thermally expanding
fireball.  It also incorporates nuclear shadowing by reducing the
input charm cross section by a factor of up to 1/3, with a centrality
dependence as estimated in \cite{Tuchin:2007pf}.  There is a fair
agreement between our \ptavg\ results and the model calculation, while
the \ptsqraa\ is not described by this prediction.  Our \ptavg\ and
\ptsqraa\ results are also compared with the calculations by Zhou et
al. \cite{Zhou:2014kka}.  These calculations are also based on a
transport approach and incorporate dissociation and regeneration of
\jpsi\ and heavier charmonia, as well as nuclear shadowing according
to EKS98 \cite{Eskola:1998df}.  While the most central data point is
matched by the prediction, it does not describe the evolution of
\ptsqraa\ towards peripheral collisions.  It must be noted that our
results from Pb-Pb collisions at forward rapidity \cite{Adams:2015aa}
exhibit a continuous decrease of \ptavg\ and \ptsqraa\ from peripheral
towards central events and are thus closer to the theory predictions,
while the behaviour of mid-rapidity Pb-Pb results is more compatible
with a flat \npart~dependence.

%
\begin{figure}
\begin{center}
\includegraphics[width=0.75\linewidth]{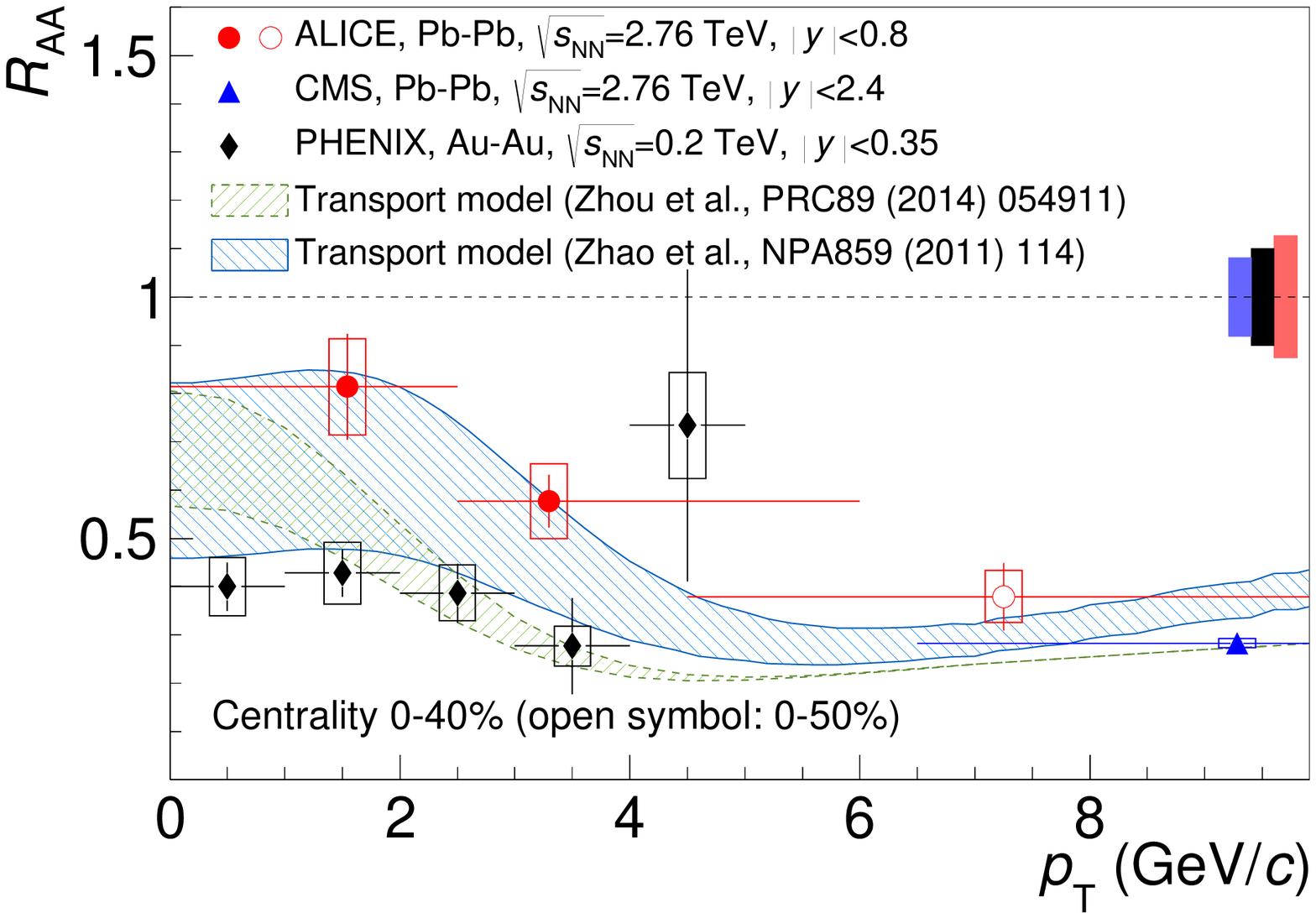}
\end{center}
\caption{\label{fig:raaptincl}
The nuclear modification factor \raa\ of inclusive \jpsi, measured at
mid-rapidity ($|y| < 0.8$) in Pb-Pb collisions ($0$--$40$\% most
central) at \sqrtsnn~= 2.76~TeV, as a function of transverse momentum
\pt.  The filled symbols are placed at the measured \pt\ for the given
interval.  Since for the data point in $4.5 < \pt < 10$~\gevc\ (open
symbol, $0$--$50$\% most central) \ptavg\ is not available due to the
limited statistics, it is plotted at the centre of the \pt~interval.
The uncorrelated systematic uncertainties (type II) are depicted by
the open boxes, while the correlated uncertainties (type I) are shown
as the filled boxes at unity.
The data are compared to corresponding results by PHENIX for Au-Au
collisions ($0$--$40$\% most central) at \sqrtsnn~= 0.2~TeV
\cite{Adare:2006ns}, by CMS for Pb-Pb collisions ($0$--$40$\% most
central) at \sqrtsnn~= 2.76~TeV \cite{Chatrchyan:2012np}, and to
predictions by the model of Zhou et al. \cite{Zhou:2014kka} and Zhao
et al \cite{Zhao:2010nk,Zhao:2011cv}.}
\end{figure}
%

The \raa\ of inclusive \jpsi\ in three \pt~intervals is shown in
\Fi{fig:raaptincl} along with the results by the CMS collaboration for
the interval $6.5 < \pt < 30$~\gevc\ \cite{Chatrchyan:2012np}, both in
$0$--$40$\% most central Pb-Pb collisions.  The corresponding numerical
values are $0.82 \pm 0.11\textrm{(stat.)} \pm 0.10\textrm{(syst.)}$
for the interval $0 < \pt < 2.5$~\gevc\ and $0.58 \pm
0.06\textrm{(stat.)} \pm 0.08\textrm{(syst.)}$ for $2.5 < \pt <
6$~\gevc, where the systematic uncertainties quoted here are the
uncorrelated (type II) ones, as listed in \Ta{tab:systUncRaa}.  The
data point for $4.5 < \pt < 10$~\gevc\ corresponds to the \raa~value
given in \Ta{tab:NonPrompt} (centrality range $0$--$50$\%).
Table~\ref{tab:NonPrompt} also contains the \raa\ values for prompt
\jpsi, which are numerically identical to the ones for inclusive
\jpsi.  The inclusive \raa\ values below $\pt = 6$~\gevc\ are
significantly higher than those measured at higher \pt, corresponding
to a decrease of \raa\ with increasing \pt, while the high~\pt\ data
point is close to the CMS measurement.  This \pt~dependence is similar
to the one observed at forward rapidity \cite{Abelev:2013ila}, and is
in clear contrast to the \pt~dependence measured at lower
centre-of-mass energies by the PHENIX collaboration for \sqrtsnn~=
0.2~TeV \cite{Adare:2006ns}.  Figure~\ref{fig:raaptincl} also shows
the model predictions by Zhou et al. \cite{Zhou:2014kka}.  The value
of the predicted \raa\ is systematically below the measurement and
exhibits a \pt~dependence similar to the one in the data.  The
prediction by Zhao et al. \cite{Zhao:2010nk,Zhao:2011cv} is close to
our result.  In both models, the rise of \raa\ towards $\pt = 0$ is
due to the dominant contribution from \jpsi\ regeneration via
coalescence.

%
\begin{figure}
\begin{center}
\begin{minipage}[b]{0.49\linewidth}
\includegraphics[width=1.0\linewidth]{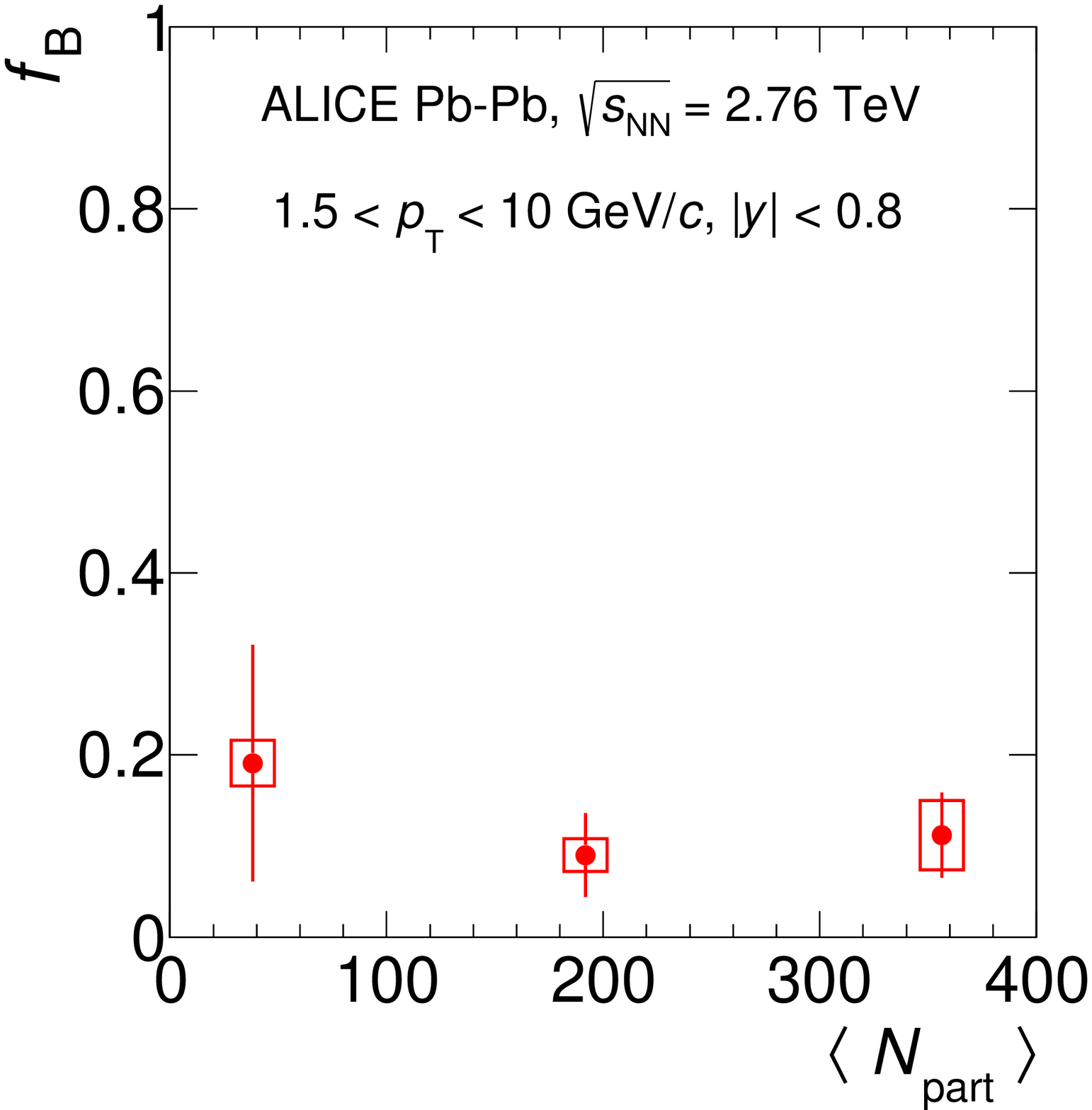}
\end{minipage}
\begin{minipage}[b]{0.49\linewidth}
\includegraphics[width=1.0\linewidth]{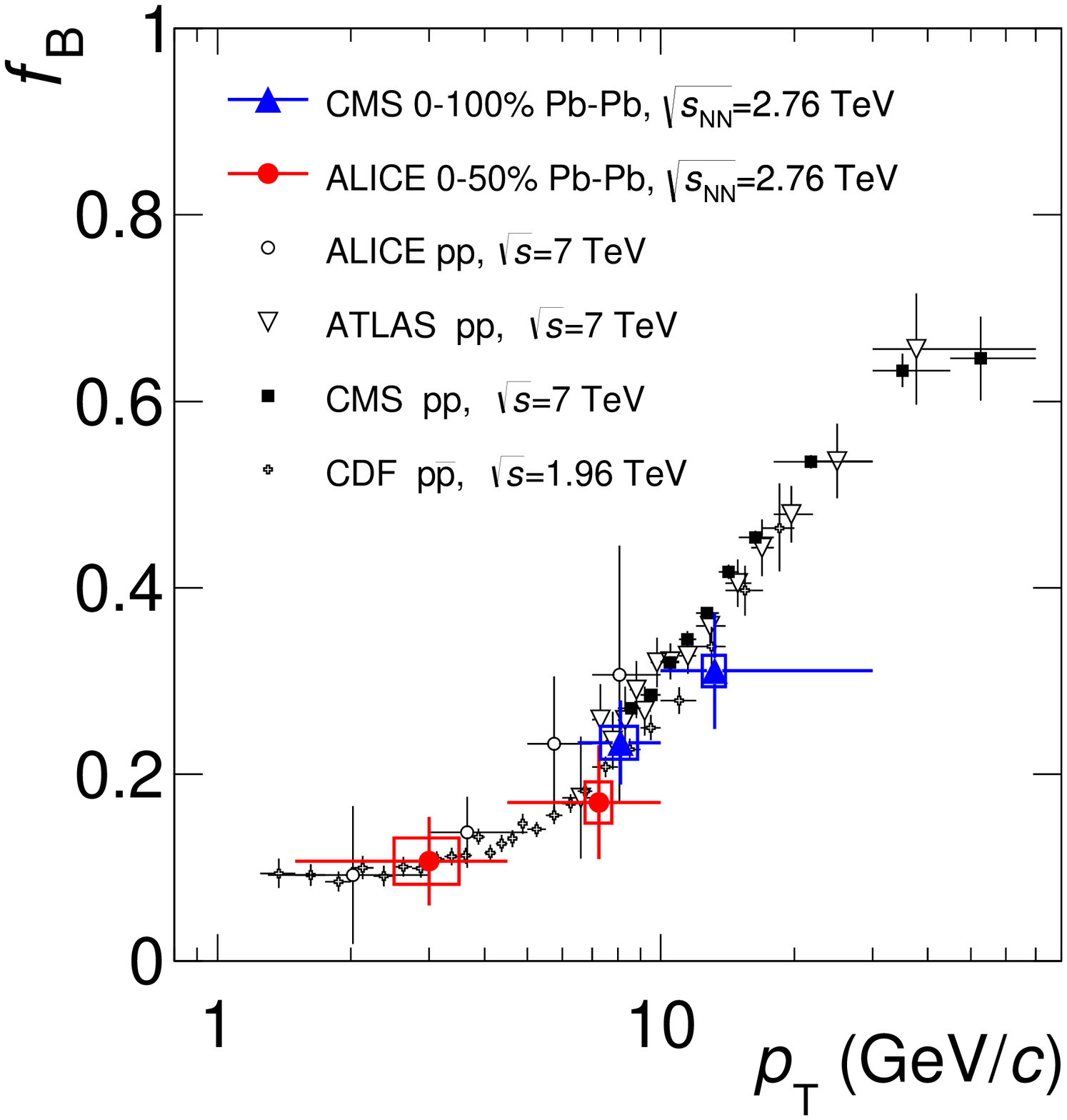}
\end{minipage}
\end{center}
\caption{\label{fig:fb}
The fraction of \jpsi\ from beauty hadron decays \fb\ at mid-rapidity
measured in the \pt~interval $1.5 < \pt < 10$~\gevc\ for centrality
selected Pb-Pb collisions at \sqrtsnn~= 2.76~TeV (left).
The \pt~dependence of \fb\ at mid-rapidity for Pb-Pb (\sqrtsnn~=
2.76~TeV, $|y_{\jpsi}| < 0.8$) and pp (\sqrts~= 7~TeV, $|y_{\jpsi}| <
0.9$) \cite{Abelev:2012gx} collisions is compared with measurements by
CDF ($|y_{\jpsi}| < 0.6$) \cite{Acosta:2004yw}, ATLAS ($|y_{\jpsi}| <
0.75$) \cite{Aad:2011sp}, and CMS ($|y_{\jpsi}| < 0.9$)
\cite{Chatrchyan:2012np,Khachatryan:2010yr} (right).}
\end{figure}
%

%
\begin{table}[ht]
\centering
\resizebox{1.0\columnwidth}{!}{
\begin{tabular}{|c|c|c|c|c|}
\hline
\pt(\gevc)  & \fb(\%)              & \raa(inclusive \jpsi)  & \raa(prompt \jpsi)     & \raa(non-prompt \jpsi) \\[1.01ex]
\hline
0.0~--~1.5  & --                   & 0.89$\pm$0.20$\pm$0.21 & --                     & --                     \\[1.01ex]
1.5~--~4.5  & 10.7$\pm$4.8$\pm$2.5 & 0.76$\pm$0.09$\pm$0.08 & 0.76$\pm$0.10$\pm$0.08 & 0.73$\pm$0.34$\pm$0.20 \\[1.01ex]
4.5~--~10.0 & 17.0$\pm$6.1$\pm$2.2 & 0.38$\pm$0.07$\pm$0.06 & 0.38$\pm$0.07$\pm$0.06 & 0.37$\pm$0.15$\pm$0.09 \\[1.01ex]
\hline
\end{tabular}
}
\caption{
The numerical values on the fraction of \jpsi\ from beauty hadron
decays \fb\ at mid-rapidity and the nuclear modification factors \raa\
of inclusive, prompt and non-prompt \jpsi\ for Pb-Pb collisions at
\sqrts~= 2.76~TeV.  These results correspond to the centrality
interval $0$--$50$\%.  The first uncertainty is statistical and the
second uncorrelated systematic (type II).
\label{tab:NonPrompt}
}
\end{table}
%

The fraction of non-prompt \jpsi\ in the \pt\ range 1.5~--~10 \gevc\
is shown as a function of the number of participants for the
centrality intervals $40$--$90$\% (\npart~= 38), $10$--$40$\%
(\npart~= 192), and $0$--$10$\% (\npart~= 356) in the left panel of
\Fi{fig:fb}.  Within uncertainties, no centrality dependence is
observed.  The \pt\ dependence of \fb\ (centrality: $0$--$50$\%) is
shown in the right panel of \Fi{fig:fb} and compared with the
measurements by CMS in the centrality interval $0$--$100$\% and $\pt >
6.5$~\gevc\ (for the numerical values see \Ta{tab:NonPrompt}).  Our
results at low transverse momenta extend the CMS measurements in Pb-Pb
collisions towards lower \pt.  Also shown are results at mid-rapidity
in pp at \sqrts~= 7~TeV (ALICE \cite{Abelev:2012gx}, ATLAS
\cite{Aad:2011sp} and CMS \cite{Khachatryan:2010yr}) and in \ppbar\
collisions at \sqrts~= 1.96~TeV (CDF \cite{Acosta:2004yw}).
Considering the ALICE and CMS results in Pb-Pb collisions together, a
similar \pt~dependence as in pp is observed.  However, this
similarity could be coincidental, being due to a compensation of the
medium effects on the prompt component (\jpsi\ dissociation and
recombination) and on the non-prompt part (b-quark energy loss).

%
\begin{figure}
\begin{center}
\begin{minipage}[b]{0.65\linewidth}
\includegraphics[width=0.9\linewidth]{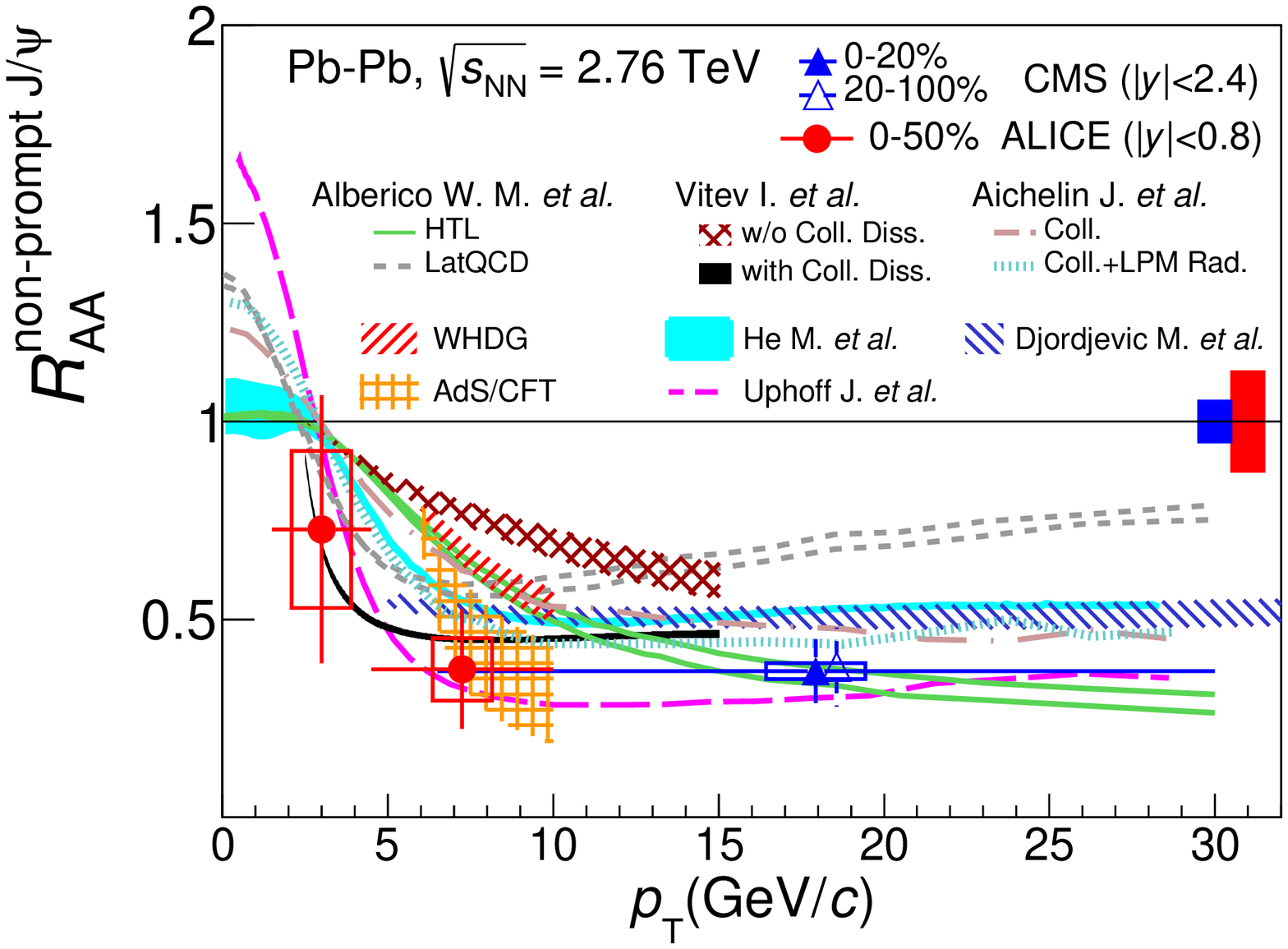}
\end{minipage}
\begin{minipage}[b]{0.34\linewidth}
\includegraphics[width=0.9\linewidth]{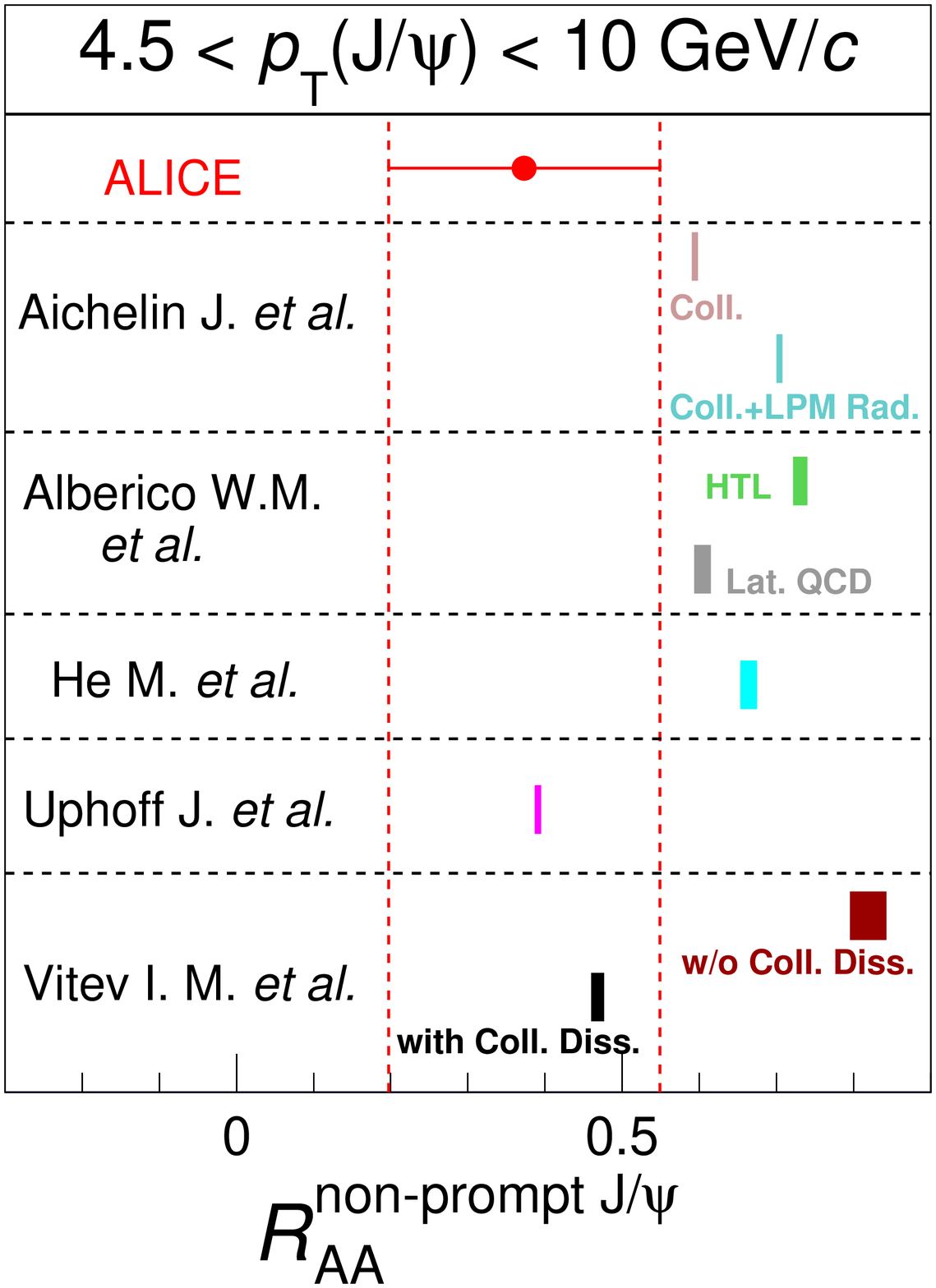}
\end{minipage}
\end{center}
\caption{
The nuclear modification factor \raa\ at mid-rapidity ($|y| < 0.8$)
for non-prompt \jpsi\ in Pb-Pb collisions at \sqrtsnn~= 2.76~TeV as a
function of transverse momentum \pt.  The ALICE measurement
corresponds to the $0$--$50$\% centrality range and to the
\pt~intervals $1.5 < \pt < 4.5$~\gevc\ and $4.5 < \pt < 10$~\gevc.
The uncorrelated systematic uncertainties (type II) are depicted by
the open boxes, while the correlated uncertainties (type I) are shown
as filled boxes at unity.
Results by CMS for higher \pt\ in the centrality range $0$--$20$\% and
$20$--$100$\% \cite{Chatrchyan:2012np} are also shown (the two points
have been slightly displaced horizontally for better visibility).  The
data are compared to theoretical predictions at mid-rapidity (see text
for details).  In the right panel, the ALICE result in the
\pt~interval $4.5 < \pt < 10$~\gevc\ is compared to theoretical
predictions integrated over the same \pt\ range.}
\label{fig:raanonprompt}
\end{figure}
%

In \Fi{fig:raanonprompt} the nuclear modification factor for
non-prompt \jpsi\ for $1.5 < \pt < 4.5$~\gevc\ and  $4.5 < \pt <
10$~\gevc\ is shown together with the result by CMS for $6.5 < \pt <
30$~\gevc\ \cite{Chatrchyan:2012np} and with theoretical model
predictions
\cite{Uphoff:2012gb,Alberico:2011zy,Alberico:2013bza,He:2014cla,Sharma:2009hn,Sharma:2012dy,Djordjevic:2013xoa,Wicks:2005gt,Aichelin:2013mra,Gossiaux:2014jga,Horowitz:2007su}.
One should note that the centrality ranges are not the same for ALICE
($0$--$50$\%) and CMS ($0$--$20$\% and $20$--$100$\%).  However, the
results obtained by CMS for these two centrality bins are compatible
with each other, and also compatible with our measurement in the high
\pt~interval ($4.5 < \pt < 10$~\gevc).  The model by Uphoff et
al.~\cite{Uphoff:2012gb} follows a partonic transport approach based
on the Boltzmann equation, which allows interactions among all
partons.  It does not include radiative processes for heavy quarks.
The calculation has been performed for a fixed impact parameter $b =
5$~fm.  In the model of Alberico et
al.~\cite{Alberico:2011zy,Alberico:2013bza} the propagation of the
heavy quarks in the medium is described by the relativistic Langevin
equation.  The predicted \pt\ dependence of \raa\ is strongly
influenced by the choice of transport coefficients.  Two values are
considered, either as provided by a perturbative calculation (hard
thermal loop approach) or extracted from lattice-QCD simulations.  The
calculations have been provided for the centrality range $0$--$50$\%.
A transport approach, which is based on a strong-coupling scheme, is
employed in the model of He et al.~\cite{He:2014cla}.  The transport
is implemented using non-perturbative interactions for heavy quarks
and mesons through the QGP, hadronization and hadronic phases of a
nuclear collision.  In particular, the elastic heavy-quark scattering
in the QGP is evaluated within a thermodynamic T-matrix approach, by
generating resonances close to the critical temperature that can in
turn recombine into B~mesons, followed by hadronic diffusion using
effective hadronic scattering amplitudes.  The hydrodynamic evolution
of the system is quantitatively constrained by the measured transverse
momentum distributions and elliptic flow of light hadrons.  Radiative
processes, which should improve the description at high \pt, are not
included in this approach.  The calculations have been performed in
the centrality range $0$--$50$\%.  The model of Vitev et
al.~\cite{Sharma:2009hn,Sharma:2012dy} assumes the existence of open
heavy flavour bound-state solutions in the QGP in the vicinity of the
critical temperature.  A description of beauty quark quenching is
combined with B~meson inelastic breakup processes.  Furthermore,
modified beauty parton distribution functions and beauty fragmentation
functions in a co-moving plasma are implemented in this calculation.
The prediction is shown for a fixed centrality, corresponding to
$\npart = 200$, a value very close to the average number of
participants in the centrality range $0$--$50$\%.  In the model, a
sizable fraction of the suppression is ascribed to the inelastic
break-up processes (collisional dissociation), as can be deduced from
\Fi{fig:raanonprompt} by comparing the full model prediction with and
without the contribution of this specific process.  The model of
Djordjevic \cite{Djordjevic:2013xoa}, shown in \Fi{fig:raanonprompt}
for the centrality range $0$--$50$\%,  uses a formalism that takes into
account finite size dynamical QCD medium with finite magnetic mass
effects and running coupling.  In the WHDG model \cite{Wicks:2005gt}
(centrality range $0$--$50$\%) the energy loss is computed using
perturbative QCD and considering both elastic and inelastic partonic
collisions and path length fluctuations.  The approach of Aichelin et
al. \cite{Aichelin:2013mra,Gossiaux:2014jga} includes a contribution
of radiative gluon emission in the interaction of heavy quarks with
light quarks, which are considered as dynamical scattering centers.
In this model the relative contribution to the energy loss by
radiative processes, as compared to collisional ones, is influenced by
introducing a finite gluon mass.  The results of the model shown in
\Fi{fig:raanonprompt}, which are obtained for the centrality range
$0$--$50$\%, correspond to either a pure collisional scenario or a
combination of collisional and radiative energy loss.  Finally, in the
model of Horowitz and Gyulassy \cite{Horowitz:2007su}, also applied to
the centrality interval $0$--$50$\%, the string inspired AdS/CFT
gravity-gauge theory correspondence
\cite{Maldacena:1997re,Gubser:1998bc} is applied to the case of heavy
quark energy loss.  In the right hand inset of \Fi{fig:raanonprompt},
the ALICE \raa\ value, integrated over the range $4.5 < \pt <
10$~\gevc, is compared to theoretical predictions computed in the same
\pt\ range.  Most of the models predict a larger value of \raa\ than
observed in the measurement.  However, more precise data are needed to
discriminate among the different models.  The next LHC run will
provide increased statistics for this measurement.

%
\section{Conclusions}

A study of \jpsi\ production at mid-rapidity in Pb-Pb collisions at
\sqrtsnn~= 2.76~TeV has been presented.  A reduction of the inclusive
\jpsi\ \ptavg\ is observed in Pb-Pb collisions in comparison to pp.
The ratio $\ptsqraa = \ptavgsq_{\rbt{AA}} / \ptavgsq_{\rbt{pp}}$ is
found to be significantly below unity, corresponding to a
medium-induced change in the shape of the \pt~spectra.  The nuclear
modification factor \raa\ depends on \pt.  It is around 0.8 for $\pt <
2.5$~\gevc\ and reaches, at higher \pt, almost the same level of
suppression as observed at RHIC energies at low \pt.  These
observations might be indicative of a sizable contribution of charm
quark coalescence to the \jpsi\ production at low \pt.  Transport
models including this additional component are able to qualitatively
describe the features seen in the data.

The fraction of \jpsi\ from beauty hadron decays is determined as a
function of centrality and \pt.  No significant centrality dependence
is observed.  By combining this measurement with the inclusive \jpsi\
results the \raa\ of non-prompt \jpsi\ is obtained in the region $1.5
< \pt\ < 10$~\gevc, thus extending the coverage of CMS to the low \pt\
region.  The nuclear modification in the region $4.5 < \pt <
10$~\gevc\ is found to be stronger than predicted by most of the
models.

%
\newenvironment{acknowledgement}{\relax}{\relax}
\begin{acknowledgement}
\section*{Acknowledgements}

The ALICE Collaboration would like to thank all its engineers and technicians for their invaluable contributions to the construction of the experiment and the CERN accelerator teams for the outstanding performance of the LHC complex.
The ALICE Collaboration gratefully acknowledges the resources and support provided by all Grid centres and the Worldwide LHC Computing Grid (WLCG) collaboration.
The ALICE Collaboration acknowledges the following funding agencies for their support in building and
running the ALICE detector:
State Committee of Science,  World Federation of Scientists (WFS)
and Swiss Fonds Kidagan, Armenia,
Conselho Nacional de Desenvolvimento Cient\'{\i}fico e Tecnol\'{o}gico (CNPq), Financiadora de Estudos e Projetos (FINEP),
Funda\c{c}\~{a}o de Amparo \`{a} Pesquisa do Estado de S\~{a}o Paulo (FAPESP);
National Natural Science Foundation of China (NSFC), the Chinese Ministry of Education (CMOE)
and the Ministry of Science and Technology of China (MSTC);
Ministry of Education and Youth of the Czech Republic;
Danish Natural Science Research Council, the Carlsberg Foundation and the Danish National Research Foundation;
The European Research Council under the European Community's Seventh Framework Programme;
Helsinki Institute of Physics and the Academy of Finland;
French CNRS-IN2P3, the `Region Pays de Loire', `Region Alsace', `Region Auvergne' and CEA, France;
German Bundesministerium fur Bildung, Wissenschaft, Forschung und Technologie (BMBF) and the Helmholtz Association;
General Secretariat for Research and Technology, Ministry of
Development, Greece;
Hungarian Orszagos Tudomanyos Kutatasi Alappgrammok (OTKA) and National Office for Research and Technology (NKTH);
Department of Atomic Energy and Department of Science and Technology of the Government of India;
Istituto Nazionale di Fisica Nucleare (INFN) and Centro Fermi -
Museo Storico della Fisica e Centro Studi e Ricerche "Enrico
Fermi", Italy;
MEXT Grant-in-Aid for Specially Promoted Research, Ja\-pan;
Joint Institute for Nuclear Research, Dubna;
National Research Foundation of Korea (NRF);
Consejo Nacional de Cienca y Tecnologia (CONACYT), Direccion General de Asuntos del Personal Academico(DGAPA), M\'{e}xico, :Amerique Latine Formation academique – European Commission(ALFA-EC) and the EPLANET Program
(European Particle Physics Latin American Network)
Stichting voor Fundamenteel Onderzoek der Materie (FOM) and the Nederlandse Organisatie voor Wetenschappelijk Onderzoek (NWO), Netherlands;
Research Council of Norway (NFR);
National Science Centre, Poland;
Ministry of National Education/Institute for Atomic Physics and Consiliul Naţional al Cercetării Ştiinţifice - Executive Agency for Higher Education Research Development and Innovation Funding (CNCS-UEFISCDI) - Romania;
Ministry of Education and Science of Russian Federation, Russian
Academy of Sciences, Russian Federal Agency of Atomic Energy,
Russian Federal Agency for Science and Innovations and The Russian
Foundation for Basic Research;
Ministry of Education of Slovakia;
Department of Science and Technology, South Africa;
Centro de Investigaciones Energeticas, Medioambientales y Tecnologicas (CIEMAT), E-Infrastructure shared between Europe and Latin America (EELA), Ministerio de Econom\'{i}a y Competitividad (MINECO) of Spain, Xunta de Galicia (Conseller\'{\i}a de Educaci\'{o}n),
Centro de Aplicaciones Tecnológicas y Desarrollo Nuclear (CEA\-DEN), Cubaenerg\'{\i}a, Cuba, and IAEA (International Atomic Energy Agency);
Swedish Research Council (VR) and Knut $\&$ Alice Wallenberg
Foundation (KAW);
Ukraine Ministry of Education and Science;
United Kingdom Science and Technology Facilities Council (STFC);
The United States Department of Energy, the United States National
Science Foundation, the State of Texas, and the State of Ohio;
Ministry of Science, Education and Sports of Croatia and  Unity through Knowledge Fund, Croatia.
Council of Scientific and Industrial Research (CSIR), New Delhi, India

\end{acknowledgement}

%
\bibliographystyle{utphys}
\bibliography{jpsi_pbpb_cernpre.bib}


%
\newpage

\appendix
\section{The ALICE Collaboration}
\label{app:collab}



\begingroup
\small
\begin{flushleft}
J.~Adam\Irefn{org40}\And
D.~Adamov\'{a}\Irefn{org83}\And
M.M.~Aggarwal\Irefn{org87}\And
G.~Aglieri Rinella\Irefn{org36}\And
M.~Agnello\Irefn{org111}\And
N.~Agrawal\Irefn{org48}\And
Z.~Ahammed\Irefn{org132}\And
S.U.~Ahn\Irefn{org68}\And
I.~Aimo\Irefn{org94}\textsuperscript{,}\Irefn{org111}\And
S.~Aiola\Irefn{org137}
% ...
\renewcommand\labelenumi{\textsuperscript{\theenumi}~}

\section*{Affiliation notes}
\renewcommand\theenumi{\roman{enumi}}
\begin{Authlist}
\item \Adef{0}Deceased
\item \Adef{idp5890544}{Also at: University of Kansas, Lawrence, Kansas, United States}
\end{Authlist}

\section*{Collaboration Institutes}
\renewcommand\theenumi{\arabic{enumi}~}
\begin{Authlist}

\item \Idef{org1}A.I. Alikhanyan National Science Laboratory (Yerevan Physics Institute) Foundation, Yerevan, Armenia
\item \Idef{org2}Benem\'{e}rita Universidad Aut\'{o}noma de Puebla, Puebla, Mexico
\item \Idef{org3}Bogolyubov Institute for Theoretical Physics, Kiev, Ukraine
\item \Idef{org4}Bose Institute, Department of Physics and Centre for Astroparticle Physics and Space Science (CAPSS), Kolkata, India
\item \Idef{org5}Budker Institute for Nuclear Physics, Novosibirsk, Russia
\item \Idef{org6}California Polytechnic State University, San Luis Obispo, California, United States
\item \Idef{org7}Central China Normal University, Wuhan, China
\item \Idef{org8}Centre de Calcul de l'IN2P3, Villeurbanne, France
\item \Idef{org9}Centro de Aplicaciones Tecnol\'{o}gicas y Desarrollo Nuclear (CEADEN), Havana, Cuba
\item \Idef{org10}Centro de Investigaciones Energ\'{e}ticas Medioambientales y Tecnol\'{o}gicas (CIEMAT), Madrid, Spain
\item \Idef{org11}Centro de Investigaci\'{o}n y de Estudios Avanzados (CINVESTAV), Mexico City and M\'{e}rida, Mexico
\item \Idef{org12}Centro Fermi - Museo Storico della Fisica e Centro Studi e Ricerche ``Enrico Fermi'', Rome, Italy
\item \Idef{org13}Chicago State University, Chicago, Illinois, USA
\item \Idef{org14}China Institute of Atomic Energy, Beijing, China
\item \Idef{org15}Commissariat \`{a} l'Energie Atomique, IRFU, Saclay, France
\item \Idef{org16}COMSATS Institute of Information Technology (CIIT), Islamabad, Pakistan
\item \Idef{org17}Departamento de F\'{\i}sica de Part\'{\i}culas and IGFAE, Universidad de Santiago de Compostela, Santiago de Compostela, Spain
\item \Idef{org18}Department of Physics and Technology, University of Bergen, Bergen, Norway
\item \Idef{org19}Department of Physics, Aligarh Muslim University, Aligarh, India
\item \Idef{org20}Department of Physics, Ohio State University, Columbus, Ohio, United States
\item \Idef{org21}Department of Physics, Sejong University, Seoul, South Korea
\item \Idef{org22}Department of Physics, University of Oslo, Oslo, Norway
\item \Idef{org23}Dipartimento di Elettrotecnica ed Elettronica del Politecnico, Bari, Italy
\item \Idef{org24}Dipartimento di Fisica dell'Universit\`{a} 'La Sapienza' and Sezione INFN Rome, Italy
\item \Idef{org25}Dipartimento di Fisica dell'Universit\`{a} and Sezione INFN, Cagliari, Italy
\item \Idef{org26}Dipartimento di Fisica dell'Universit\`{a} and Sezione INFN, Trieste, Italy
\item \Idef{org27}Dipartimento di Fisica dell'Universit\`{a} and Sezione INFN, Turin, Italy
\item \Idef{org28}Dipartimento di Fisica e Astronomia dell'Universit\`{a} and Sezione INFN, Bologna, Italy
\item \Idef{org29}Dipartimento di Fisica e Astronomia dell'Universit\`{a} and Sezione INFN, Catania, Italy
\item \Idef{org30}Dipartimento di Fisica e Astronomia dell'Universit\`{a} and Sezione INFN, Padova, Italy
\item \Idef{org31}Dipartimento di Fisica `E.R.~Caianiello' dell'Universit\`{a} and Gruppo Collegato INFN, Salerno, Italy
\item \Idef{org32}Dipartimento di Scienze e Innovazione Tecnologica dell'Universit\`{a} del  Piemonte Orientale and Gruppo Collegato INFN, Alessandria, Italy
\item \Idef{org33}Dipartimento Interateneo di Fisica `M.~Merlin' and Sezione INFN, Bari, Italy
\item \Idef{org34}Division of Experimental High Energy Physics, University of Lund, Lund, Sweden
\item \Idef{org35}Eberhard Karls Universit\"{a}t T\"{u}bingen, T\"{u}bingen, Germany
\item \Idef{org36}European Organization for Nuclear Research (CERN), Geneva, Switzerland
\item \Idef{org37}Excellence Cluster Universe, Technische Universit\"{a}t M\"{u}nchen, Munich, Germany
\item \Idef{org38}Faculty of Engineering, Bergen University College, Bergen, Norway
\item \Idef{org39}Faculty of Mathematics, Physics and Informatics, Comenius University, Bratislava, Slovakia
\item \Idef{org40}Faculty of Nuclear Sciences and Physical Engineering, Czech Technical University in Prague, Prague, Czech Republic
\item \Idef{org41}Faculty of Science, P.J.~\v{S}af\'{a}rik University, Ko\v{s}ice, Slovakia
\item \Idef{org42}Faculty of Technology, Buskerud and Vestfold University College, Vestfold, Norway
\item \Idef{org43}Frankfurt Institute for Advanced Studies, Johann Wolfgang Goethe-Universit\"{a}t Frankfurt, Frankfurt, Germany
\item \Idef{org44}Gangneung-Wonju National University, Gangneung, South Korea
\item \Idef{org45}Gauhati University, Department of Physics, Guwahati, India
\item \Idef{org46}Helsinki Institute of Physics (HIP), Helsinki, Finland
\item \Idef{org47}Hiroshima University, Hiroshima, Japan
\item \Idef{org48}Indian Institute of Technology Bombay (IIT), Mumbai, India
\item \Idef{org49}Indian Institute of Technology Indore, Indore (IITI), India
\item \Idef{org50}Inha University, Incheon, South Korea
\item \Idef{org51}Institut de Physique Nucl\'eaire d'Orsay (IPNO), Universit\'e Paris-Sud, CNRS-IN2P3, Orsay, France
\item \Idef{org52}Institut f\"{u}r Informatik, Johann Wolfgang Goethe-Universit\"{a}t Frankfurt, Frankfurt, Germany
\item \Idef{org53}Institut f\"{u}r Kernphysik, Johann Wolfgang Goethe-Universit\"{a}t Frankfurt, Frankfurt, Germany
\item \Idef{org54}Institut f\"{u}r Kernphysik, Westf\"{a}lische Wilhelms-Universit\"{a}t M\"{u}nster, M\"{u}nster, Germany
\item \Idef{org55}Institut Pluridisciplinaire Hubert Curien (IPHC), Universit\'{e} de Strasbourg, CNRS-IN2P3, Strasbourg, France
\item \Idef{org56}Institute for Nuclear Research, Academy of Sciences, Moscow, Russia
\item \Idef{org57}Institute for Subatomic Physics of Utrecht University, Utrecht, Netherlands
\item \Idef{org58}Institute for Theoretical and Experimental Physics, Moscow, Russia
\item \Idef{org59}Institute of Experimental Physics, Slovak Academy of Sciences, Ko\v{s}ice, Slovakia
\item \Idef{org60}Institute of Physics, Academy of Sciences of the Czech Republic, Prague, Czech Republic
\item \Idef{org61}Institute of Physics, Bhubaneswar, India
\item \Idef{org62}Institute of Space Science (ISS), Bucharest, Romania
\item \Idef{org63}Instituto de Ciencias Nucleares, Universidad Nacional Aut\'{o}noma de M\'{e}xico, Mexico City, Mexico
\item \Idef{org64}Instituto de F\'{\i}sica, Universidad Nacional Aut\'{o}noma de M\'{e}xico, Mexico City, Mexico
\item \Idef{org65}iThemba LABS, National Research Foundation, Somerset West, South Africa
\item \Idef{org66}Joint Institute for Nuclear Research (JINR), Dubna, Russia
\item \Idef{org67}Konkuk University, Seoul, South Korea
\item \Idef{org68}Korea Institute of Science and Technology Information, Daejeon, South Korea
\item \Idef{org69}KTO Karatay University, Konya, Turkey
\item \Idef{org70}Laboratoire de Physique Corpusculaire (LPC), Clermont Universit\'{e}, Universit\'{e} Blaise Pascal, CNRS--IN2P3, Clermont-Ferrand, France
\item \Idef{org71}Laboratoire de Physique Subatomique et de Cosmologie, Universit\'{e} Grenoble-Alpes, CNRS-IN2P3, Grenoble, France
\item \Idef{org72}Laboratori Nazionali di Frascati, INFN, Frascati, Italy
\item \Idef{org73}Laboratori Nazionali di Legnaro, INFN, Legnaro, Italy
\item \Idef{org74}Lawrence Berkeley National Laboratory, Berkeley, California, United States
\item \Idef{org75}Lawrence Livermore National Laboratory, Livermore, California, United States
\item \Idef{org76}Moscow Engineering Physics Institute, Moscow, Russia
\item \Idef{org77}National Centre for Nuclear Studies, Warsaw, Poland
\item \Idef{org78}National Institute for Physics and Nuclear Engineering, Bucharest, Romania
\item \Idef{org79}National Institute of Science Education and Research, Bhubaneswar, India
\item \Idef{org80}Niels Bohr Institute, University of Copenhagen, Copenhagen, Denmark
\item \Idef{org81}Nikhef, National Institute for Subatomic Physics, Amsterdam, Netherlands
\item \Idef{org82}Nuclear Physics Group, STFC Daresbury Laboratory, Daresbury, United Kingdom
\item \Idef{org83}Nuclear Physics Institute, Academy of Sciences of the Czech Republic, \v{R}e\v{z} u Prahy, Czech Republic
\item \Idef{org84}Oak Ridge National Laboratory, Oak Ridge, Tennessee, United States
\item \Idef{org85}Petersburg Nuclear Physics Institute, Gatchina, Russia
\item \Idef{org86}Physics Department, Creighton University, Omaha, Nebraska, United States
\item \Idef{org87}Physics Department, Panjab University, Chandigarh, India
\item \Idef{org88}Physics Department, University of Athens, Athens, Greece
\item \Idef{org89}Physics Department, University of Cape Town, Cape Town, South Africa
\item \Idef{org90}Physics Department, University of Jammu, Jammu, India
\item \Idef{org91}Physics Department, University of Rajasthan, Jaipur, India
\item \Idef{org92}Physik Department, Technische Universit\"{a}t M\"{u}nchen, Munich, Germany
\item \Idef{org93}Physikalisches Institut, Ruprecht-Karls-Universit\"{a}t Heidelberg, Heidelberg, Germany
\item \Idef{org94}Politecnico di Torino, Turin, Italy
\item \Idef{org95}Purdue University, West Lafayette, Indiana, United States
\item \Idef{org96}Pusan National University, Pusan, South Korea
\item \Idef{org97}Research Division and ExtreMe Matter Institute EMMI, GSI Helmholtzzentrum f\"ur Schwerionenforschung, Darmstadt, Germany
\item \Idef{org98}Rudjer Bo\v{s}kovi\'{c} Institute, Zagreb, Croatia
\item \Idef{org99}Russian Federal Nuclear Center (VNIIEF), Sarov, Russia
\item \Idef{org100}Russian Research Centre Kurchatov Institute, Moscow, Russia
\item \Idef{org101}Saha Institute of Nuclear Physics, Kolkata, India
\item \Idef{org102}School of Physics and Astronomy, University of Birmingham, Birmingham, United Kingdom
\item \Idef{org103}Secci\'{o}n F\'{\i}sica, Departamento de Ciencias, Pontificia Universidad Cat\'{o}lica del Per\'{u}, Lima, Peru
\item \Idef{org104}Sezione INFN, Bari, Italy
\item \Idef{org105}Sezione INFN, Bologna, Italy
\item \Idef{org106}Sezione INFN, Cagliari, Italy
\item \Idef{org107}Sezione INFN, Catania, Italy
\item \Idef{org108}Sezione INFN, Padova, Italy
\item \Idef{org109}Sezione INFN, Rome, Italy
\item \Idef{org110}Sezione INFN, Trieste, Italy
\item \Idef{org111}Sezione INFN, Turin, Italy
\item \Idef{org112}SSC IHEP of NRC Kurchatov institute, Protvino, Russia
\item \Idef{org113}SUBATECH, Ecole des Mines de Nantes, Universit\'{e} de Nantes, CNRS-IN2P3, Nantes, France
\item \Idef{org114}Suranaree University of Technology, Nakhon Ratchasima, Thailand
\item \Idef{org115}Technical University of Ko\v{s}ice, Ko\v{s}ice, Slovakia
\item \Idef{org116}Technical University of Split FESB, Split, Croatia
\item \Idef{org117}The Henryk Niewodniczanski Institute of Nuclear Physics, Polish Academy of Sciences, Cracow, Poland
\item \Idef{org118}The University of Texas at Austin, Physics Department, Austin, Texas, USA
\item \Idef{org119}Universidad Aut\'{o}noma de Sinaloa, Culiac\'{a}n, Mexico
\item \Idef{org120}Universidade de S\~{a}o Paulo (USP), S\~{a}o Paulo, Brazil
\item \Idef{org121}Universidade Estadual de Campinas (UNICAMP), Campinas, Brazil
\item \Idef{org122}University of Houston, Houston, Texas, United States
\item \Idef{org123}University of Jyv\"{a}skyl\"{a}, Jyv\"{a}skyl\"{a}, Finland
\item \Idef{org124}University of Liverpool, Liverpool, United Kingdom
\item \Idef{org125}University of Tennessee, Knoxville, Tennessee, United States
\item \Idef{org126}University of the Witwatersrand, Johannesburg, South Africa
\item \Idef{org127}University of Tokyo, Tokyo, Japan
\item \Idef{org128}University of Tsukuba, Tsukuba, Japan
\item \Idef{org129}University of Zagreb, Zagreb, Croatia
\item \Idef{org130}Universit\'{e} de Lyon, Universit\'{e} Lyon 1, CNRS/IN2P3, IPN-Lyon, Villeurbanne, France
\item \Idef{org131}V.~Fock Institute for Physics, St. Petersburg State University, St. Petersburg, Russia
\item \Idef{org132}Variable Energy Cyclotron Centre, Kolkata, India
\item \Idef{org133}Vin\v{c}a Institute of Nuclear Sciences, Belgrade, Serbia
\item \Idef{org134}Warsaw University of Technology, Warsaw, Poland
\item \Idef{org135}Wayne State University, Detroit, Michigan, United States
\item \Idef{org136}Wigner Research Centre for Physics, Hungarian Academy of Sciences, Budapest, Hungary
\item \Idef{org137}Yale University, New Haven, Connecticut, United States
\item \Idef{org138}Yonsei University, Seoul, South Korea
\item \Idef{org139}Zentrum f\"{u}r Technologietransfer und Telekommunikation (ZTT), Fachhochschule Worms, Worms, Germany
\end{Authlist}
\endgroup

\end{document}